\documentclass{aa}  

\usepackage{graphicx}

\usepackage{txfonts}
\usepackage{hyperref}
\usepackage{placeins}

\usepackage{defs}

\begin{document}

\def\mstar{M_{\rm S}}
\def\rstar{R_{\rm S}}
\def\mplanet{M_{\rm P}}
\def\rplanet{R_{\rm P}}
\def\logpeff{\log(p_{\rm eff})}
\def\spitzer{\textsc{Spitzer}}
\def\tess{\textsc{TESS}}
\def\plato{\textsc{PLATO}}
\def\cheops{\textsc{CHEOPS}}
\def\hst{\textsc{HST}}
\def\hstwfc{\textsc{HST/WFC3\_IR~G141}}
\def\hstgfourthreezero{\textsc{HST/G430L}}
\def\hstgsevenfivezero{\textsc{HST/G750L}}
\def\jwst{\textsc{JWST}}
\def\miri{\textsc{JWST/MIRI}}
\def\nircam{JWST/NIRCam}
\def\nirspec{JWST/NIRSpec}
\def\spitzeriracthree{\textsc{Spitzer/IRAC}~3.6~$\mu$m}
\def\spitzeriracfour{\textsc{Spitzer/IRAC}~4.5~$\mu$m}
\def\spitzeriracfive{\textsc{Spitzer/IRAC}~5.6~$\mu$m}
\def\spitzeriracseven{\textsc{Spitzer/IRAC}~7.6~$\mu$m}
\def\spitzerirsred{\textsc{Spitzer/IRS~Red}~22~$\mu$m}
\def\spitzerirsblue{\textsc{Spitzer/IRS~Blue}~15~$\mu$m}
\def\spitzermipstwentyfour{\textsc{Spitzer/MIPS}~24~$\mu$m}
\def\criresplus{CRIRES$+$}
\def\cgs{$CGS$~units}

\title{Phase-curve approach to study atmospheric flows in hot Jupiters}

\author{D. Shulyak\inst{1}
       \and
       W. Dietrich\inst{2}
       \and
       V. Parmentier\inst{3}
       \and
       D. Cont\inst{4,5}
       \and
       L.-M. Lara\inst{1}
       \and
       L. Gkouvelis\inst{1}
       \and
       M. R. Swain\inst{6}
       \and
       M. Rengel\inst{2}
       }

\institute{
Instituto de Astrof\'{\i}sica de Andaluc\'{\i}a - CSIC, c/ Glorieta de la Astronom\'{\i}a s/n, 18008 Granada, Spain\\
\email{shulyak@iaa.es}
\and
Max-Planck Institut f\"ur Sonnensystemforschung, Justus-von-Liebig-Weg 3, D-37077, G\"ottingen, Germany
\and
Laboratoire Lagrange, Observatoire de la C\^ote d'Azur, CNRS, Universit\'e C\^ote d'Azur, Nice, France
\and
Universit\"ats-Sternwarte, Ludwig-Maximilians-Universit\"at, M\"unchen, Scheinerstrasse 1, 81679 M\"unchen, Germany
\and
Exzellenzcluster Origins, Boltzmannstrasse 2, 85748 Garching bei München, Germany
\and
California Institute of Technology, NASA Jet Propulsion Laboratory, USA
          }

\date{Received ; accepted}

 
\abstract
 {Short-orbit gaseous exoplanets are the best targets to study atmospheric dynamics. Their hot atmospheres
 generate emission that is strong enough to be detected and analyzed in detail.
 A time series of such emission observations collected at various photometric filters~--~phase curves~--~provides insights into atmospheric flows. 
 Modern observations reveal a wide variety of phase curves, but their utility for probing atmospheric circulation as a function of altitude 
 has not yet been explored in detail.}
 {We aim to understand the properties of phase curves and their connection to underlying atmospheric flows, as well as 
 to define a set of multiwavelength observations that could be used to resolve these flows as a function of altitude.}
 {We utilized a {subset of the solar metallicity models from the} grid of hot Jupiters calculated by ADAM (former SPARC/MITgcm) to predict phase curves.
 To do this we used state-of-the-art radiative transfer codes and the latest opacity sources. 
 We made predictions for a variety of photometric filters on board the \spitzer, \tess, \cheops, \hst, and \jwst\ missions, and explored
 the sensitivity of each filter to flows at various atmospheric depths.}
 {Our calculations show that the main parameter that regulates the phase-curve offsets in our models is the atmospheric temperature, 
 where hotter planets tend to have smaller offsets, {although high metallicity can also have strong impact by reducing phase-curve offsets.}
 This is not fully supported by available observations, which possibly indicates a missing physical process in the models.
 The predicted contribution functions suggest that the best combination of photometric filters 
 to study atmospheric flows is \nircam\ filters because they are sensitive to a wide range of pressures 
 between 10~bar and 10$^{-4}$~bar depending on planet temperature, respectively.
 High-resolution spectroscopy is predicted to detect differential Doppler shifts of $\Delta\upsilon$$\approx$1~-~4~\kms\ between molecular bands 
 formed at different altitudes, providing an independent probe of vertical circulations.}
 {}


\maketitle

\nolinenumbers

\section{Introduction}

Modern advances in observations of exoplanet atmospheres have helped us
to better understand their properties and evolution pathways. We test our knowledge 
by building theoretical models and comparing them to the data. One of the challenges
in exoplanet science is to understand the structure and atmospheric dynamics of a class
of hot exoplanets called hot Jupiters (HJs). These are gas giant planets unlike any in our Solar System. 
They orbit very close to their host stars, typically with periods of only a few days, and receive intense stellar radiation 
that heats their atmospheres to thousands of kelvin \citep{2021A&A...653A..52F,2015ApJ...799..229W,2010ApJ...718L.145W}. 
Because their atmospheres are extremely hot, these planets emit enough light to be detected from Earth.

Hot Jupiters orbit their host stars in synchronous rotation such that the dayside is permanently irradiated, 
while the opposite nightside is colder. 
The heat absorbed at the dayside is redistributed by atmospheric circulations to the nightside. 
The exact flow pattern is a complicated function of atmospheric density, spectral
type of the host star, and orbital distance \citep{2025A&A...699A..74A}. 
Often, an equatorial jet in the direction of rotation can be detected 
\citep[see, e.g.,][for the most extreme case of the WASP-127~b and references therein]{2025A&A...693A.213N}. 
If winds are strong and deep enough where the gas density is high, the energy absorbed at the substellar point 
can be re-emitted by a moving gas at a different location \citep[e.g.,][]{2018haex.bookE.116P}.
Thus, the maximum of the observed emission from the planet will no longer coincide
with the rotation phase at which the substellar point {lies along the observer's line of sight}, but will
be shifted in a prograde (eastward) or retrograde (westward) direction.
Thus, analysis of exoplanet flux variation with rotation phase~--~phase curves~--~is a powerful 
method to learn about atmospheric winds. Two properties of phase curves are usually investigated:
phase-curve offset (measures the displacement of the light maximum relative to the substellar point) 
and phase-curve amplitude (measures the relative flux difference between day and night sides, respectively).
Moreover, looking at different wavelengths can help 
to resolve the vertical structure of winds due to the variation in atmospheric opacity.

Multiwavelength observations of phase curves with, for example, the past \spitzer\ mission (Spitzer Space Telescope) equipped
with several broadband photometric filters revealed a rather complicated picture of phase-curve offsets
in a variety of hot exoplanets. Both prograde and retrograde offsets were observed and attempts were made to understand them
using theoretical models \citep[see, e.g.,][and references therein]{2018haex.bookE.116P}.
A typical challenge was that the observed phase shifts did not exhibit a strong correlation with any specific atmospheric parameter.
For instance, \citet{2018haex.bookE.116P} found a weakly constrained
anti-correlation of the phase-curve offsets with planet equilibrium temperature, $\teq$. Another study by \citet{2022AJ....163..256M}
also states a similar weakly constrained trend of decreasing offsets as $\teq$ increases.

General circulation models (GCM) are usually used to model atmospheric dynamics. 
This became a classical approach \citep{2002A&A...385..166S} 
and has been used ever since at various levels of complexity \citep[see, e.g.,][]{2025arXiv250921588S,Showman2020}. 
Importantly, GCMs can capture the horizontal heat transport, which is directly probed by the phase curves. 
These simulations predict a decrease in the phase-curve offset
as atmospheric temperature rises \citep{2024MNRAS.531.1056R,2016ApJ...828...22P,2013ApJ...776..134P}, 
while new models that include different approaches
for cloud formation indicate that this trend may be reversed in some cases \citep{2021ApJ...908..101R}.
In addition, full 3D, anelastic models show that both phase-curve offsets are possible \citep{2025ApJ...995...84D}. 
Additionally, recently developed models that include magnetic dynamos indicate that the planetary magnetic field
can reverse the atmospheric flows to produce retrograde phase-curve offsets \citep{2025A&A...699A.339B,2025ApJ...978..149H,2017NatAs...1E.131R}.
Despite these advances, a systematic framework linking multiwavelength phase-curve observables to atmospheric circulation properties is still lacking.

One of the big challenges is the lack of accurate and sufficient measurements of the phase curves of exoplanets with
diverse properties. Currently operating \tess\ (Transiting Exoplanet Survey Satellite), \cheops\ (CHaracterising ExOPlanet Satellite),
and  \textsc{HST} (Hubble Space Telescope)
missions offer only a few photometric wavelengths
in visual and near-infrared. A breakthrough is expected from \jwst\ (James Webb Space Telescope) observations due to the numerous photometric filters
offered for phase-curve observations.
This motivates us to test state-of-the-art models of atmospheric circulation against present and future observations, as well as to investigate their
application to study atmospheric dynamics from multiwavelength observations.

In this work, we test whether multiband phase curves can uniquely diagnose vertical wind structure and identify 
the origin of the model–observation discrepancy in hot Jupiter phase offsets. 
We further discuss and compare our findings with selected observations, 
and propose photometric filters that are best suited for the observations of atmospheric flows.

\section{Methods}\label{sec:methods}

\subsection{SPARC/MITgcm models}

In this work, we calculate phase curves using grid of GCM presented in \citet{2024MNRAS.531.1056R}. 
The grid was calculated by means of the non-gray SPARC/MITgcm code described in \citet{2009ApJ...699..564S}
and is publicly available\footnote{https://zenodo.org/doi/10.5281/zenodo.10785320}\footnote{https://3dsim.oca.eu/hot-jupiters-3d-model}.
The original grid spawns a wide range
of equilibrium temperatures between $\teq$$=$1000~K and $\teq$$=$2400~K. Individual models
also assume a range in metallicity between $\mh$$=$+0.0 and $\mh$$=$1.5 and rotation periods between $P$$=$0.27~d and $P$$=$23.13~d, respectively.
All models with $\teq$$<$1400~K were computed by artificially excluding \tio/\vo\ absorption to mimic the condensation of these molecules
at cold temperatures, while hotter models were constructed both with and without \tio/\vo.
Although \citet{2024MNRAS.531.1056R} already provided the emission flux and phase curves from each of the model 
in a number of photometric bands, in this work we calculate it again. 
We do this because we now include more complete up-to-date opacity sources
and calculate contribution functions to estimate the light formation depths in all considered photometric bands that were additionally
extended to include some of \jwst\ instruments.
We also calculate high-resolution spectra to investigate the velocity shifts of spectral lines caused by atmospheric flows.

In our study, we test the sensitivity of individual photometric filters
to the atmospheric flows. We therefore only consider a subset of models that have solar metallicity $\mh$$=$+0.0 and infinite drag time. 
Including models with additional parameter value (e.g., metallicity) would make our calculations difficult to 
complete in reasonable time. Also, we limit ourselves to models that include \tio/\vo\ for planets with $\teq$$\geqslant$1400~K. 
Following the original grid, the planets are assumed to have fixed radius $\rplanet$$=$1.3~$\Rjup$, but various masses 
$\mplanet$$=$0.46~$\Mjup$ ($\logg$$=$2.8~cgs), $\mplanet$$=$1.36~$\Mjup$ ($\logg$$=$3.3~cgs),
and $\mplanet$$=$4.3~$\Mjup$ ($\logg$$=$3.8~cgs), respectively.
More details can be found in the original paper by \citet{2024MNRAS.531.1056R}.

\subsection{Predicting planet emission}

In order to predict radiative flux from the planet
we used a modified version of the \taurex~(Tau Retrieval for Exoplanets) 
forward model \citep{2015ApJ...802..107W,2015ApJ...813...13W}. In particular, a major modification
was made to include more realistic radiative transfer solvers that account
for the scattering contribution in the source function. 
In this work, we used the linear Feautrier form of the radiative transfer \citep{1978stat.book.....M} as a compromise 
between computational speed and accuracy. All numerical routines were taken from the \llmodels\ stellar
atmosphere code \citep{2004A&A...428..993S}.

The molecular line lists are taken from the \textsc{DACE}\footnote{\tt https://dace.unige.ch/dashboard/} online database
which used the \heliosk\ code\footnote{\tt https://github.com/exoclime/HELIOS-K} 
to generate molecular opacity tables \citep{2015ApJ...808..182G,2021ApJS..253...30G}.
We also included atomic line opacity due to elements C, O, Na, Mg, Si, K, Ca, Ti, Cr, and Fe using up-to-date line lists 
by R.~Kurucz\footnote{\tt http://kurucz.harvard.edu} \citep{2018ASPC..515...47K}.

The continuum opacity includes
Rayleigh scattering on molecules and collisionally induced absorption 
due to H$_{\rm 2}$-H$_{\rm 2}$ and H$_{\rm 2}$-He either
after \citet{2011mss..confEFC07A,2012JChPh.136d4319A} or \citet{2001JQSRT..68..235B,2002A&A...390..779B,1989ApJ...341..549B}, respectively.
Additional opacity sources relevant for hot atmospheres such as  bound-free and free-free transitions of \hminus, 
Rayleigh scattering on \ion{H}{i} atoms, and Thomson scattering on free electrons, are taken into
account as described in \citet{2020A&A...639A..48S}.

The calculation of emergent flux was carried out assuming equilibrium chemistry and local thermodynamic equilibrium (LTE).
The first assumption means that we ignore mixing processes (e.g., turbulent transport, molecular diffusion), and photochemistry. 
Note that we do consider removal of atmospheric constituents by condensation.
The second assumption, LTE, means that the population of energy levels within atoms and molecules,
as well as the velocity distribution of free electrons, obey Boltzmann-Maxwell distribution. 

\subsection{Phase-curve modeling}

To generate a phase curve we used the original grid resolution of published SPARC/MITgcm models which 
consists of 64 longitude and 32 latitude points, respectively (i.e., constant 5.625\degr\ resolution). 
The number of original pressure levels is 53, but it is increased to 100 when using \taurex. 
For each pixel, we calculate equilibrium concentrations (considering condensation) of atmospheric species using
\textsc{FastChem}\footnote{\tt https://github.com/exoclime/FastChem} code \citep{2018MNRAS.479..865S}.
Then, we calculate the intensity of emergent radiation at the planet's surface for eleven angles
$\mu=\cos\theta$ between line of sight and the direction of radiation. 
The surface flux at a given orbital phase $\phi$ and photometric filter $f$ is obtained by disk integration 
of surface intensity:

\begin{equation}
    F_{\phi,f} = \displaystyle \int\limits_S I_{\mu,\phi,f} ds = \displaystyle\frac{\sum_i I_{\mu_i,\phi_i,f} s_i}{\sum_i s_i},
\end{equation}

\noindent
where $s_i$ is projected area of the $i$-th surface element and the sum runs through all surface elements
that are visible at a given rotation phase $\phi$ (the sum in the denominator is simply a normalization factor). 
$I_{\mu_{i},\phi_{i},f}$ is the intensity of radiation in the direction of the observer weighted over the filter 
passband $\Phi_{f,\lambda}$:

\begin{equation}
    I_{\mu_{i},\phi_{i},f} = \displaystyle \frac{\int I(\mu_i,\phi_i,\lambda)\Phi_{f,\lambda}d\lambda}{\int \Phi_{f,\lambda}d\lambda}.
\end{equation}

\noindent
The integration is performed for a number of orbital phases to produce a smooth phase curve of the planet. 
The inclination angle of the planet's orbit is assumed to be 90\degr\ (i.e., equator-on viewing geometry).

\subsection{Contribution function and effective depth of formation}

To understand the predicted phase shifts we calculated contribution functions at each photometric filter
in order to estimate the formation depth of the outgoing radiation. This is a powerful approach
to understand average properties of emission in HJs with complex atmospheric flows \citep[see, e.g.,][]{2017ApJ...851L..26D}.
The contribution function at a given 
wavelength and directional angle $\mu$ is defined according
to \citet{1986A&A...163..135M} but, instead of its original
formulation in terms of optical depth scale, we have rewritten it in terms of logarithm of the pressure which is a more
commonly used quantity in planetary science:

\begin{equation}
\displaystyle C_{j,\lambda,\mu}(\log p_{j}) = \mu^{-1} \ln{(10)} p_{j} \displaystyle\frac{\kappa_{j,\lambda}}{g_{j}} S_{j,\lambda} \; \displaystyle e^{-\tau_{j,\lambda}/\mu},
\label{eq:confun}
\end{equation}

\noindent
where $p_j$ is the local pressure at the $j$-th atmospheric layer,
$\kappa_{j,\lambda}$ is the total absorption coefficient (i.e., including scattering), 
$g_{j}$ is the local gravity, 
$S_{j,\lambda}$ is the source function, 
and $\tau_{j,\lambda}$ is the monochromatic optical depth, respectively. 
The source function is calculated assuming isotropic scattering:

\begin{equation}
S_{j,\lambda} = (1 - \alpha_{j,\lambda}) B_{j,\lambda} \; + \; \alpha_{j,\lambda} J_{j,\lambda} \; + \; \alpha_{j,\lambda} J_{j,\lambda}^{\rm ext},
\label{eq:source_function}
\end{equation}

\noindent
where $\alpha_{j,\lambda}$ is the single scattering albedo, $B_{j,\lambda}$ is the Planck function, and
$J_{j,\lambda}$ and $J_{j,\lambda}^{\rm ext}$ are the values of the mean intensity of the planetary and stellar radiation, respectively. 
The latter can be defined as

\begin{equation}
J_{j,\lambda}^{\rm ext}  =  \displaystyle\frac{1}{4\pi} \int\limits_{\Omega}I_{\mathrm{TOA},\lambda}^{\rm ext}(\omega) \; e^{-\tau_{j,\lambda}/\mu_{\omega}} \, d\omega = 
                            \displaystyle\frac{1}{4\pi} \; H_{\mathrm{TOA},\lambda}^{\rm ext} \; e^{-\tau_{j,\lambda}/\mu_{\rm ext}},
\label{eq:j_ext}
\end{equation}

\noindent
where $\mu_{\rm ext}$ is the angle between normal to the surface and the direction to the star,
and $H_{\mathrm{TOA},\lambda}$ is the flux of the host star entering top of the planetary atmosphere.
The stellar flux was taken from the BT-NextGen (AGSS2009) grid of models \citep{2011ASPC..448...91A,2012RSPTA.370.2765A}.
The BT-NextGen models and filter passbands are available 
from the Spanish Virtual Observatory (SVO)\footnote{https://svo.cab.inta-csic.es/main/index.php}, for example.

We calculate $C_{j,\lambda,\mu}$ for a set of fifteen $\mu$ angles and then disk integrate
for each rotation phase $\phi$ and pressure level $j$. 
These disk-integrated contribution functions, $C_{j,\lambda,\phi}$ are then used to obtain
monochromatic effective pressure for a given rotation phase ($N$ is the number of atmospheric layers)

\begin{equation}
\displaystyle \logpeff{_{\lambda,\phi}} = \frac{\sum_{j=1}^N \log(p_j)\,C_{j,\lambda,\phi}}{\sum_{j=1}^N C_{j,\lambda,\phi}},
\label{eq:peff_lambda}
\end{equation}

\noindent
and finally the effective formation depth $\logpeff$ as 
a weighted mean over the filter passband $\Phi_{f,\lambda}$:

\begin{equation}
\displaystyle \logpeff = \frac{\int \logpeff{_{\lambda,\phi}}\, \Phi_{f,\lambda}\, d\lambda}{\int \Phi_{f,\lambda}\,d\lambda}.
\label{eq:peff}
\end{equation}
\noindent
As seen from the definitions above, $\logpeff$ quantifies an effective pressure (i.e., altitude) 
where the bulk of the outgoing radiation is formed when looking at particular wavelengths (i.e., photometric filter).
It does not, however, provide an information that is easy to interpret
because of averaging over all atmospheric layers. This becomes particularly difficult 
in cases of very broad contribution functions that spans many pressure scale heights, for example,
or contribution functions with complex shapes (e.g., a double peak function).

\section{Results}

\begin{figure*}[ht!]
\centerline{
\includegraphics[width=\hsize]{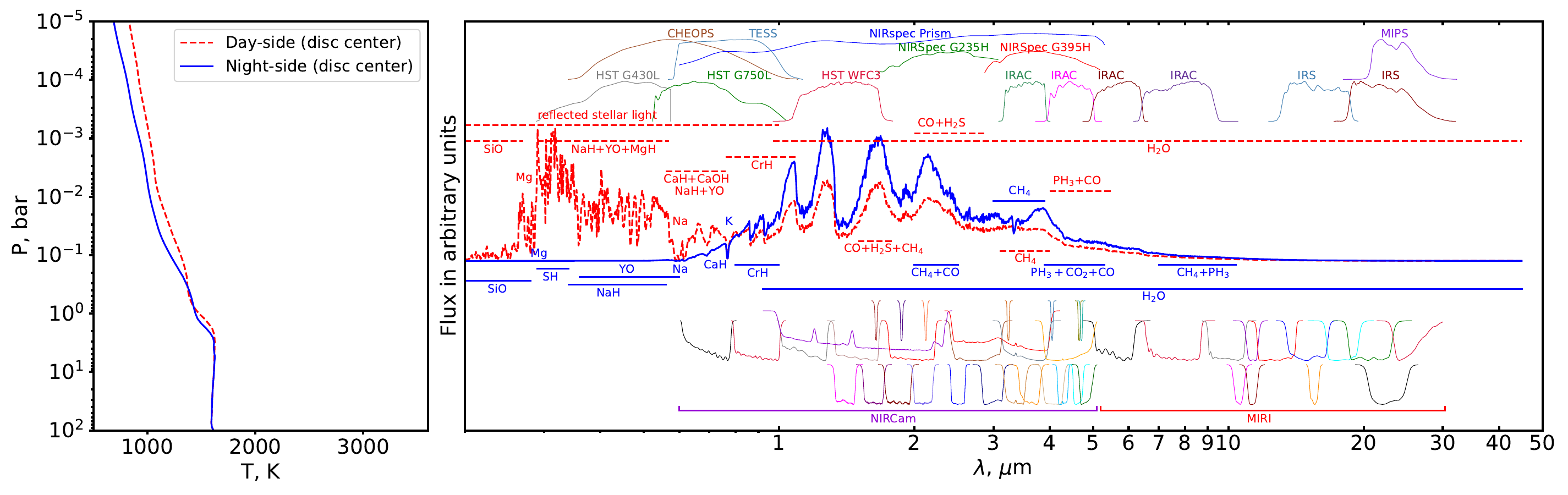}
}
\centerline{
\includegraphics[width=\hsize]{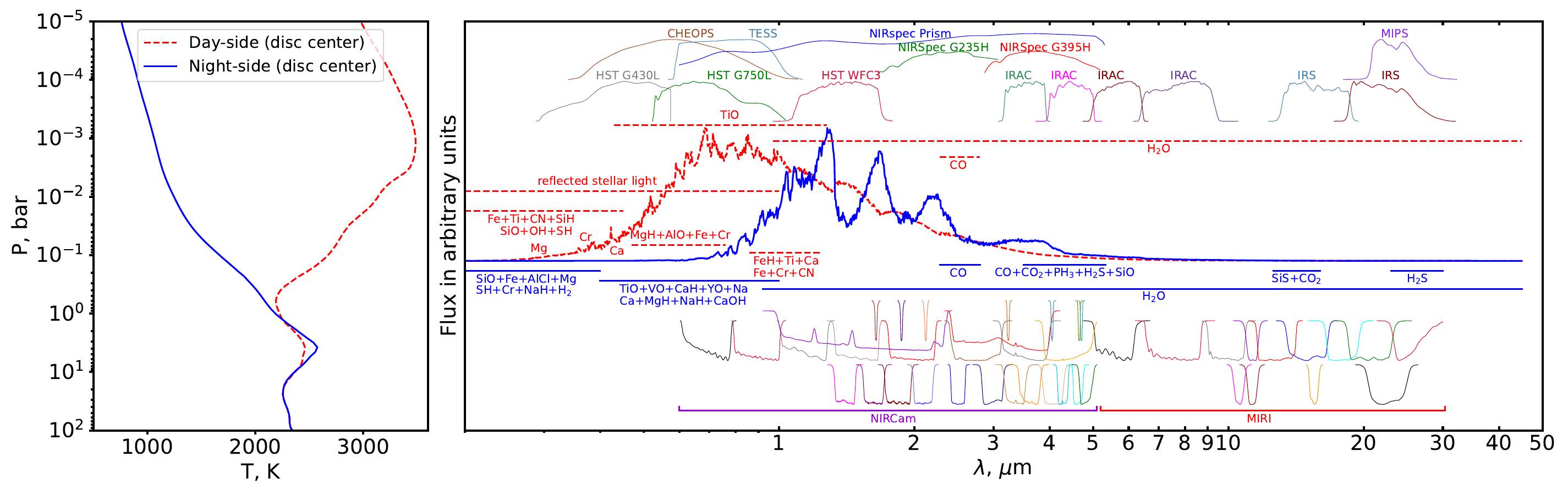}
}
\caption{Example plots of the emission spectra for planets with $\logg=3.3$~cgs 
and temperatures $\teq=1200$~K (top panel) and $\teq=2200$~K (bottom panel), respectively.
On each panel the temperature distribution (left) and emergent flux (right) at two different locations on the planet's surface are shown:
substellar point (dashed red line) and antistellar point (solid blue line). 
The response curves of photometric filters of the \spitzer, \hst, \tess, \cheops, and \jwst\ missions are shown with thin lines
of different colors. Spectral features due to prominent atmospheric contributors are marked under or above corresponding
wavelengths. The theoretical flux and filter response curves were arbitrarily scaled for a better view.}
\label{fig:t-flux-filters}
\end{figure*}

Most of the instruments that we consider in our study are equipped with medium- and broadband filters
that cover various molecular bands of different intensity. Exceptions are seven \nircam\ filters with relatively narrow
passbands, as illustrated in Fig.~\ref{fig:t-flux-filters} where we show the emission flux and locations
of prominent spectroscopic features for the two planets with $\teq$$=$1200~K and $\teq$$=$2200~K, respectively ($\logg$$=$3.3~cgs).
Many \miri\ and \spitzer\ filters are going to be affected primarily only by \hho\ lines on the planetary day and night sides. 
However, \spitzeriracthree\ and
\spitzeriracfour\ are also influenced by a few other molecules whose opacity is strongest on the nightside.
The \nircam\ filters cover strong molecular bands of \hho\ and \tio, \vo, \chhhh, \coo, and \co\ , for example, hence  allowing for a much wider range
of sensitivity to the atmospheric conditions between day and night sides. Moreover, \nircam\ contains two filters that cover
wavelengths $\lambda$$<$1~\mum, which is particularly important because short wavelengths are expected to be sensitive
to the high altitudes on the dayside if strong temperature inversion is present, due to \tio\ and/or \vo, for example. 
However, these filters can also probe deep layers 
on the nightside where the short wavelength opacity becomes weaker compared to the infrared one.
Same is true for other short wavelength filters such as \tess, \cheops, \hstgfourthreezero, and \hstgsevenfivezero, respectively. 
Importantly, on the dayside, the filters with wavelengths $\lambda$$<$1~\mum\ are affected by the stellar reflected light,
which is expected to be strong in planets with low $\teq$ because the planetary emission in this spectral region is weak.
However, the individual sensitivity of each filter can deviate from this simplistic picture due to the presence of various molecular bands 
which can have very different strength depending on local atmospheric conditions.
Below we examine the sensitivity of each filter to the atmospheric structure 
by looking at the phase-curve offsets.

\subsection{Examples of phase curves and light formation depths}

\begin{figure*}[ht!]
\centerline{
\includegraphics[width=0.33\hsize,trim=0 5.5cm 0cm 0cm,clip]{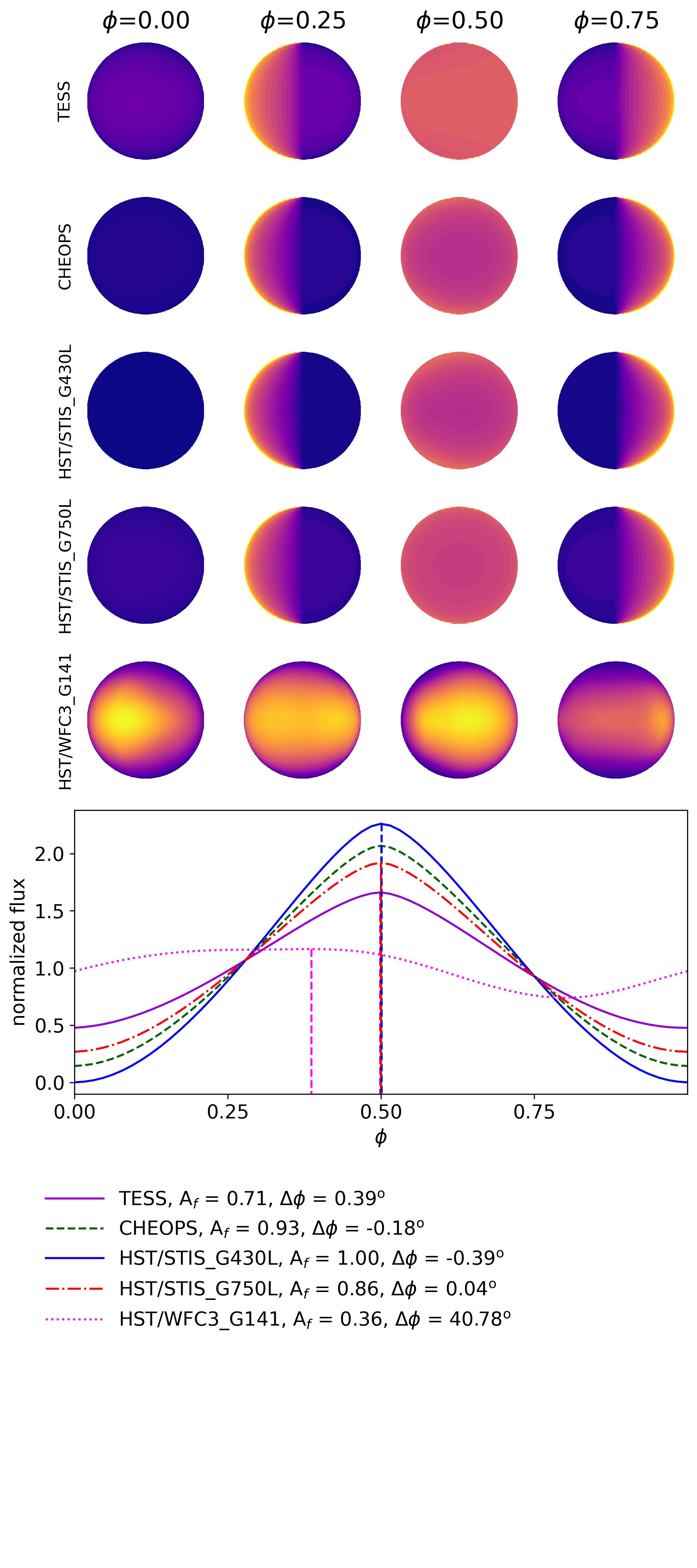}
\includegraphics[width=0.33\hsize,trim=0 5.5cm 0cm 0cm,clip]{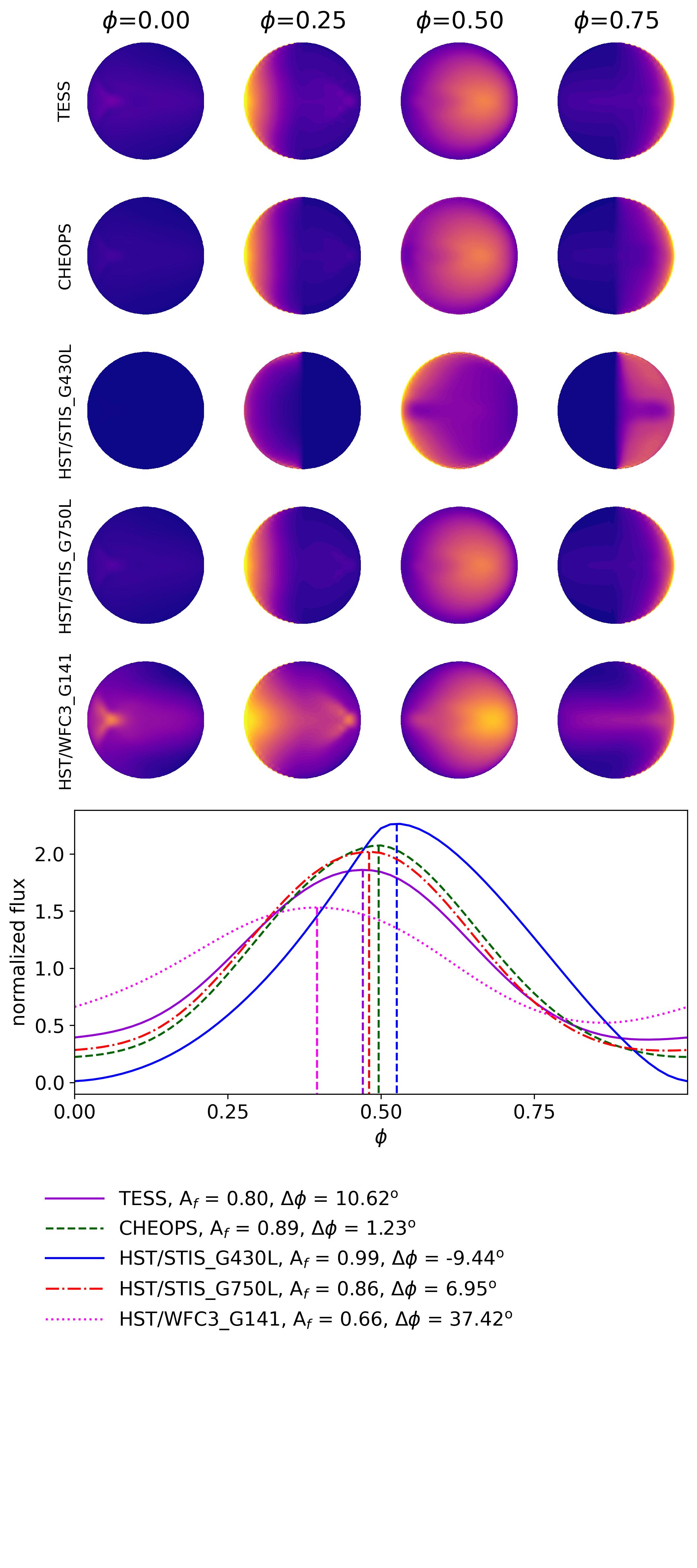}
\includegraphics[width=0.33\hsize,trim=0 5.5cm 0cm 0cm,clip]{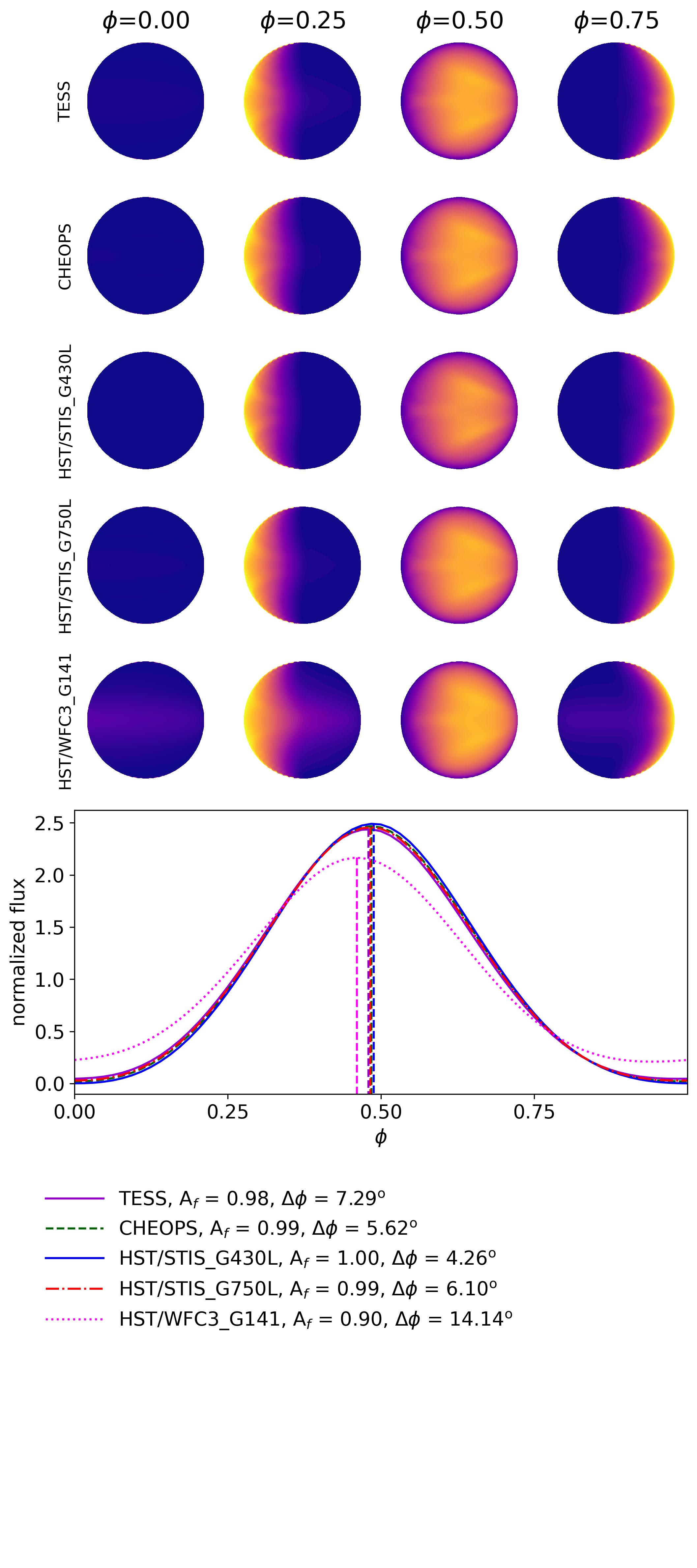}
}
\centerline{
\includegraphics[width=0.33\hsize,trim=0 0cm 0cm 0cm,clip]{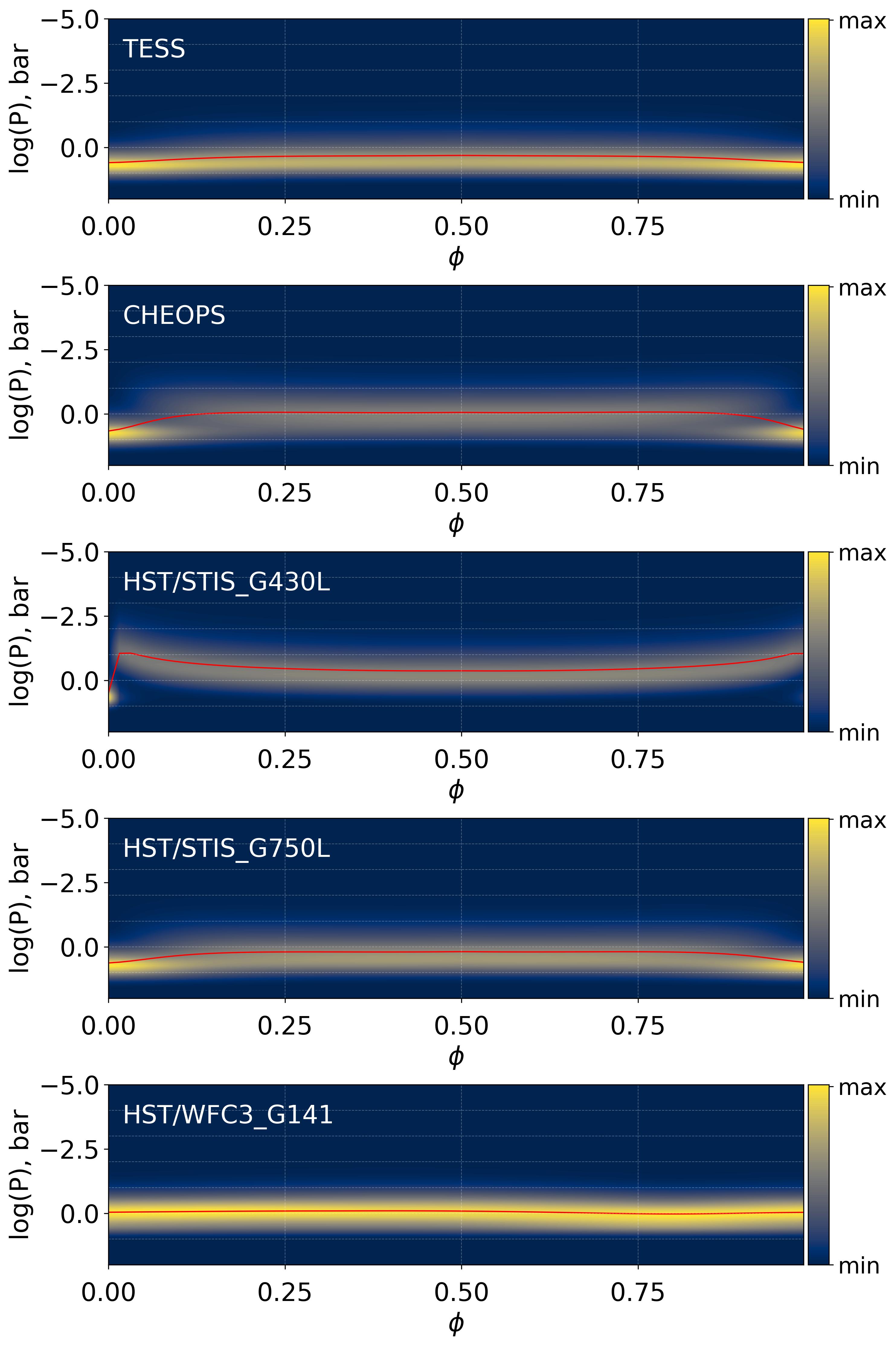}
\includegraphics[width=0.33\hsize,trim=0 0cm 0cm 0cm,clip]{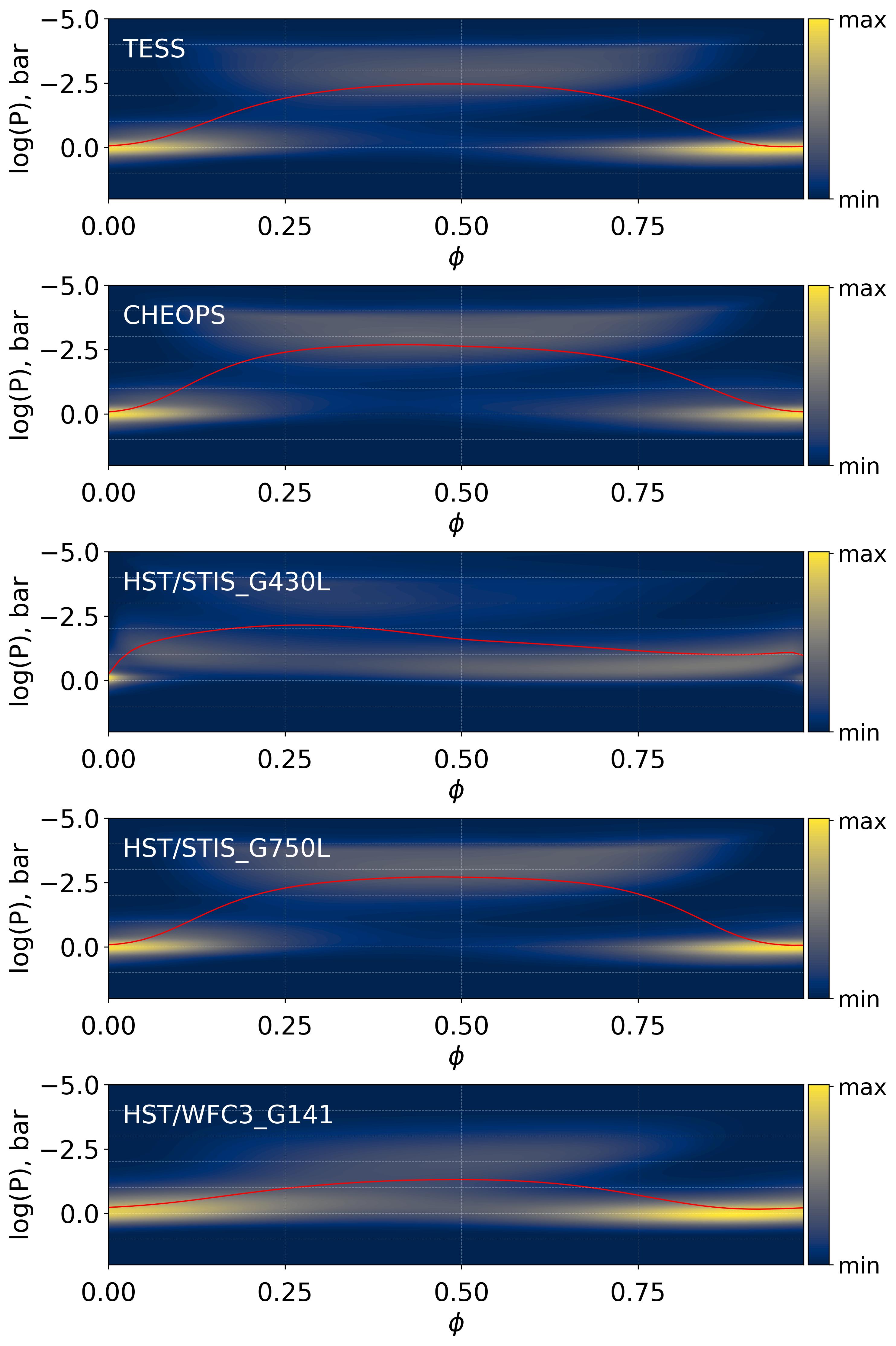}
\includegraphics[width=0.33\hsize,trim=0 0cm 0cm 0cm,clip]{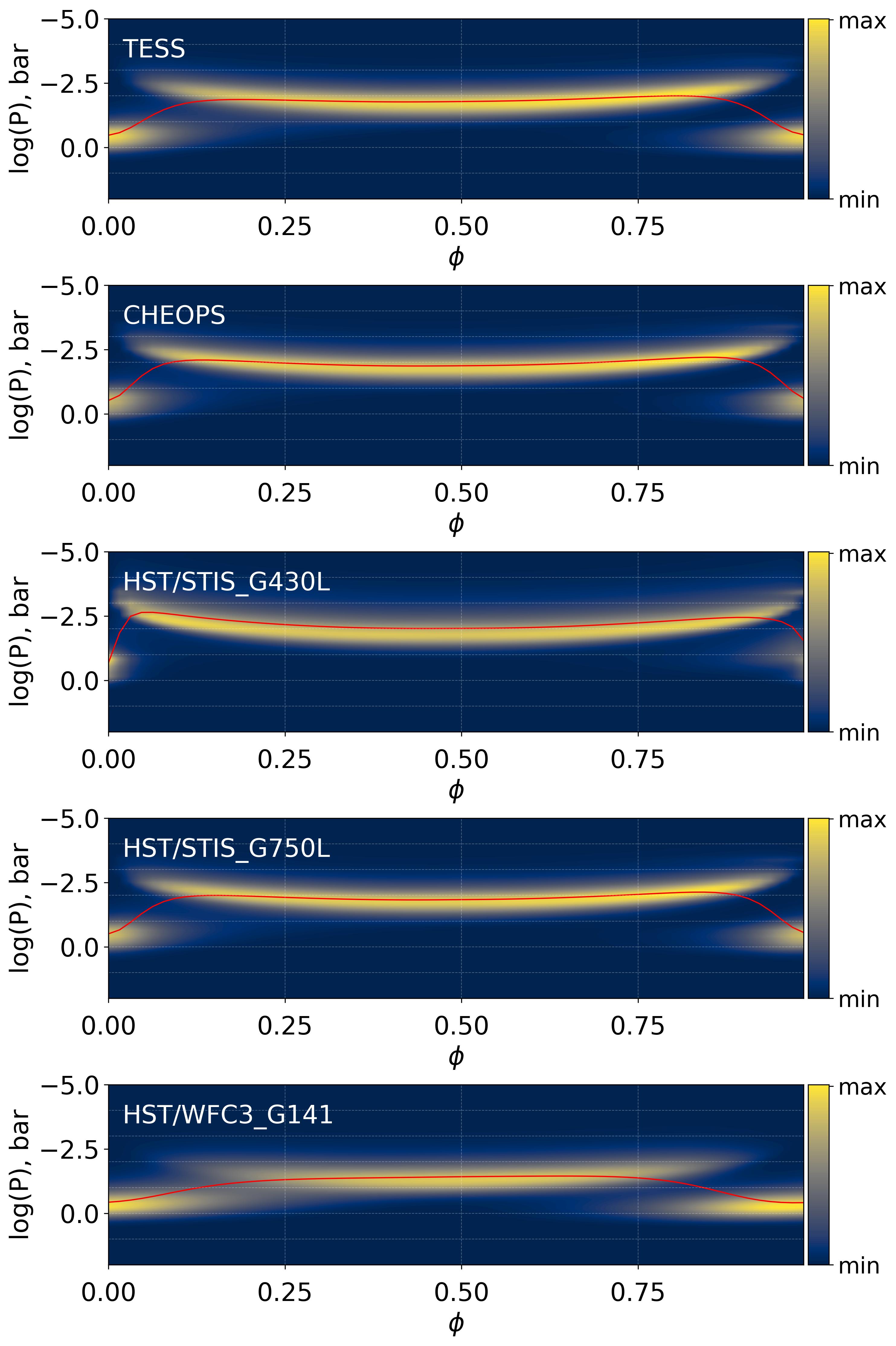}
}
\caption{Examples of synthetic phase curves calculated from three models:
$\teq$$=$1000~K (left), $\teq$$=$1600~K (middle), and $\teq$$=$2400~K (right).
The system parameters are $\mstar$$=$1.1~$\Msun$, $\mplanet$$=$1.36~$\Mjup$ ($\logg$$=$3.3~cgs).
From top to bottom: brightness images shown for the \cheops, \tess, and \hst\ filters (brightness scale is linear); 
normalized phase curves, where predicted phase shifts and phase-curve amplitudes (defined as $A_{f}$=($F_{\rm max}$ $-$ $F_{\rm min}$)/$F_{\rm max}$) 
are given in the figure's legend below the plots (locations of the maximum of the phase curves are also marked with dashed lines on the plots); 
2D maps of contribution functions for each filter
and rotation phase together with a calculated value of effective light formation depths, $\logpeff$, shown as a solid red line.}
\label{fig:lc-cf-example}
\end{figure*}

The phase curves calculated from the SPARC/MITgcm demonstrate some distinct patterns 
that could be seen from Fig.~\ref{fig:lc-cf-example} where we use examples of \cheops, \tess, and \hst\ filters.
In particular, a model with $\teq$$=$1000~K produces no significant phase-curve offsets in short wavelength filters
because most of the emission comes from the reflected stellar light which homogeneously illuminates the substellar point
and is not connected to the underlying atmospheric flows. But at longer wavelengths (e.g., \hstwfc) 
the contribution from the reflected light fades away and the emission from hot atmospheric layers dominates.
This is when significant phase-curve offsets become visible.
In planets with hotter equilibrium temperatures the light formation moves toward higher altitudes and also
the reflected light contribution decreases. This results in variability of phase-curve offsets, some of which can even
become negative (e.g., $\teq$$=$1600~K model, filter \hstgfourthreezero). 
As $\teq$ continues to increase, the offsets in all filters decrease.
Note that some filters show complex contribution functions (e.g., double peaks for the $\teq$$=$1600~K model at the day and night
terminators and nightside for the $\teq$$=$2400~K model, respectively), 
where a given filter would collect light both from high and low altitudes at some rotation phases.

\subsection{Phase-curve offsets and amplitudes}

\begin{figure*}[ht!]
\centerline{
\includegraphics[width=\hsize,trim=0cm 13.7cm 0cm    0cm,clip]{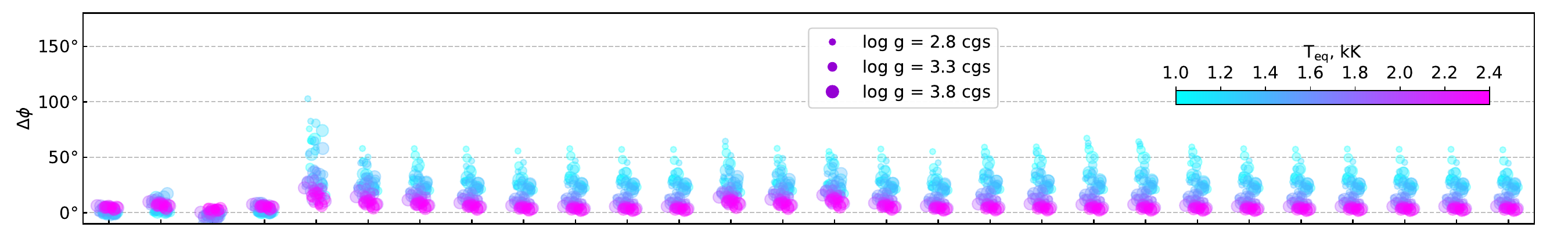}
}
\centerline{
\includegraphics[width=\hsize,trim=0cm     0cm 0cm 10.34cm,clip]{figures/offset-filter-set-1.pdf}
}
\centerline{
\includegraphics[width=\hsize,trim=0cm   12.3cm 0cm    0cm,clip]{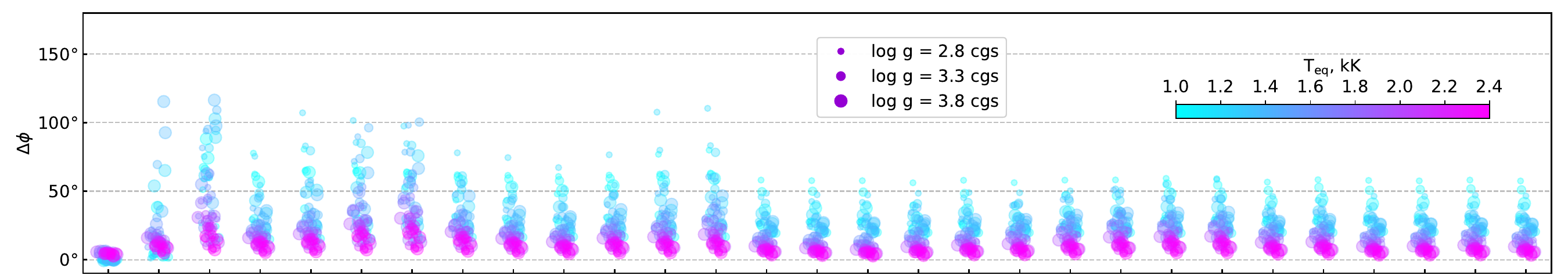}
}
\centerline{
\includegraphics[width=\hsize,trim=0cm     0cm 0cm 10.3cm,clip]{figures/offset-filter-set-2.pdf}
}
\caption{Predicted phase-curve offsets in \cheops, \tess, \hst, \spitzer, \nirspec, \miri\ (two top panels), and \nircam\ filters (two bottom panels).
The symbols are color coded according to equilibrium temperature or rotation period, and symbol size reflects the $\logg$ of the planet (see plot legends). 
The horizontal dashed line marks zero offset. 
Each point was slightly shifted along the x-axis by a random value in order to separate individual models within each filter for a better view.}
\label{fig:offset-filters-set-teq_prd}
\end{figure*}

\begin{figure*}[ht!]
\centerline{
\includegraphics[width=\hsize]{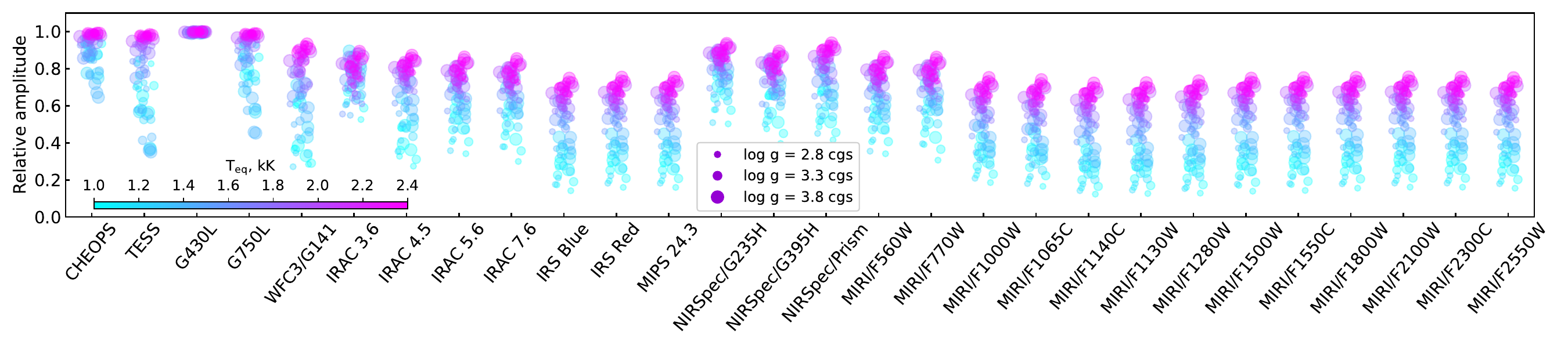}
}
\centerline{
\includegraphics[width=\hsize]{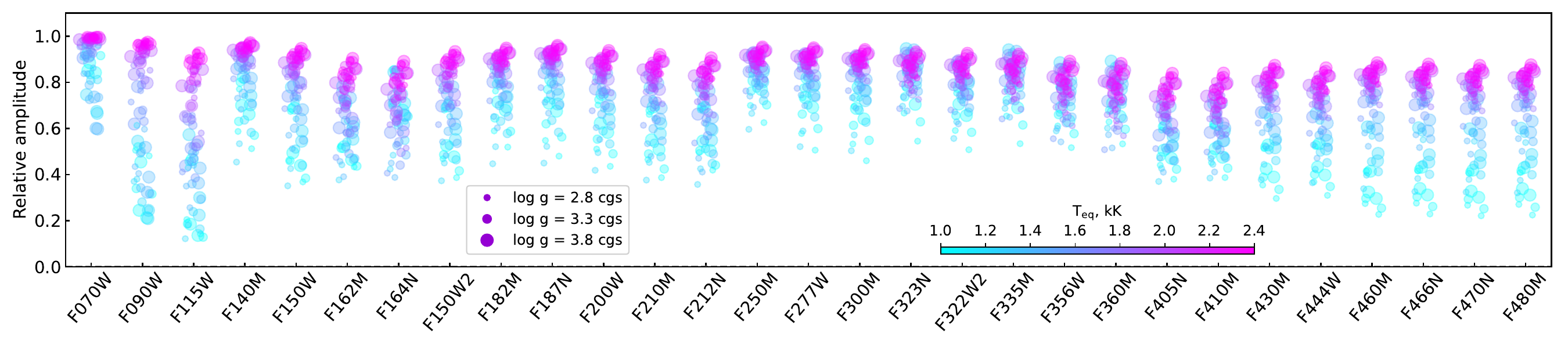}
}
\caption{Predicted phase-curve amplitudes in \cheops, \tess, \hst, \spitzer, \nirspec, \miri\ (top panel)
and \nircam\ filters (bottom panel).
The symbols are color coded according to the equilibrium temperature.
The symbol size reflects the $\logg$ of the planet. 
Each model was shifted along the x-axis by a random value in order to separate individual models within each filter for a better view.}
\label{fig:af}
\end{figure*}

We summarize predictions of the phase-curve offsets as a function of $\teq$ and rotation period in Fig.~\ref{fig:offset-filters-set-teq_prd}.
Two additional Fig.~\ref{fig:offset-filters-set-1-symlog} and Fig.~\ref{fig:offset-filters-set-2-symlog} show same
plots but in logarithm scale to visualize small offset values better.
Also, Figs.~\ref{fig:cf-tess-cheops-hst-spitzer}, \ref{fig:cf-nirspec-miri}, \ref{fig:cf-nircam1}, and ~\ref{fig:cf-nircam2}
provide averaged over day (0.25$\leqslant$$\phi$$<$0.75) and night (0.75$\leqslant$$\phi$$<$0.25) sides values of $\logpeff$ and contribution functions 
calculated from the data similar to the one shown in the bottom panels of Fig.~\ref{fig:lc-cf-example}. These figures serve to roughly represent the region 
of atmospheric depths probed by each filter depending on planetary $\teq$ and $\logg$.

The analysis of the plots reveals several features. First, it is evident that the phase-curve offset is strongly
anti-correlated with the planetary $\teq$ for all filters that cover infrared wavelengths. In filters that cover
visible spectral region and hence where the contribution from the reflected light is significant no
correlation could be seen and the phase-curve offsets are marginal
(e.g., in filters \tess, \cheops, \hstgfourthreezero, \hstgsevenfivezero, and \nircam~F070W, respectively).
The importance of the planetary temperature in regulating
the offset was noted in previous studies, too \citep[see, e.g.,][]{2002A&A...385..166S,2024MNRAS.531.1056R},
and is explained as an interplay between radiative and advective timescales in different parts of a planetary atmosphere. 
In this study we show that this is the case across multiple spectral domains.
Not surprisingly, there is a clear, though not strictly linear, dependence on rotation period as well
(see Fig.~\ref{fig:offset-filters-set-teq_prd}).
For instance, in planets orbiting low mass star $\mstar$$=$0.8~$\Msun$ larger offsets are found in longer period planets in most filters.
As the mass of the parent star increases, some filters would show larger offsets for slightly shorter rotation periods, 
especially in dense planets. We argue that this happens because, as rotation period decreases, 
the flow becomes very patchy and asymmetric at lower pressures that are probed by numerous filters.

The dependence on planetary gravity is more complex and is illustrated in 
Figs.~\ref{fig:offset-teq-logg-ms0.8}, \ref{fig:offset-teq-logg-ms1.1}, and \ref{fig:offset-teq-logg-ms1.5}.
Here we mention some prominent features. 
For the models with $\teq$$\geqslant$1600~K the gravity plays 
only a marginal role across all filters showing the maximum change in the phase-curve offset between models with smallest and largest gravity
$\delta\phi_{\rm \logg}$$=$23$\degr$ in the \nircam~F115W filter and only for the $\teq$$=$1600~K, $\mstar$$=$1.5$\Msun$ model,
while for the rest filters and models $\delta\phi_{\rm \logg}$$\leqslant$10$\degr$.
As $\teq$ increases, this difference drops below 3$\degr$. 
For models with $\teq$$\leqslant$1400~K the impact of gravity is the largest and the planet rotation rate introduces
additional constrain. In particular, in planets with the longest rotation period ($\mstar$$=$1.5$\Msun$),
the gravity impact is very strong in filters \nircam\ F090W ($\delta\phi_{\rm \logg}$$=$112$\degr$),
F162M ($\delta\phi_{\rm \logg}$$=$68$\degr$), and F164N ($\delta\phi_{\rm \logg}$$=$71$\degr$) for $\teq$$=$1400~K models.
But as $\teq$ decreases, the gravity impact weakens and is around $\delta\phi_{\rm \logg}$$\approx$80$\degr$
in filters \hstwfc, \nircam\ F212N, F210M, F164N, F162M, F150W, and F115W for the $\teq$$=$1200~K model
and then drops sharply to $\delta\phi_{\rm \logg}$$\leqslant$10$\degr$ in several filters and $\teq$$=$1000~K model, respectively.
As planetary rotation period decreases, the gravity impact decreases, too.
For planets orbiting $\mstar$$=$1.1$\Msun$, the gravity impact results in maximum $\delta\phi_{\rm \logg}$$=$71$\degr$
in \nircam\ F090W for the $\teq$$=$1400~K, decreasing to $\delta\phi_{\rm \logg}$$\approx$63$\degr$ when $\teq$$=$1200~K
with several more filters showing $\delta\phi_{\rm \logg}$$\approx$40$\degr$ (e.g., \nircam\ F212M, F200W, F210M, F140M, and F182M),
and further decreases to $\delta\phi_{\rm \logg}$$\approx$50$\degr$ for the $\teq$$=$1000~K model and filters such as
\nircam\ F212N, F210M, F150W, F140M, and \hstwfc. Finally, for planets with shortest rotation periods ($\mstar$$=$0.8$\Msun$)
the trend with gravity is slightly different. Here, the magnitude of gravity impact of about 
$\delta\phi_{\rm \logg}$$=$37$\degr$ in filter \nircam\ F090W and $\teq$$=$1200~K model, 
and, on the average, is $\delta\phi_{\rm \logg}$$\approx$25$\degr$ in all \spitzer\ filters,
\nirspec\ G395H, \miri\ F560W, F770W, F2100W, F2300C, F2550W, and \nircam\ F090W, F250M, F277W, F300M, F323N, F322W2,
F335M, F356W, and F360M, respectively.
Lastly, we note that there is no strict rule how each filter reacts to the changes in planet mass because atmospheric flows
strongly depend on rotation period. In general, we see that filters that cover short wavelengths $\lambda<1$~\mum\ demonstrate, on average,
an increase in phase-curve offset as gravity increases for some values of $\teq$, while infrared filters, again on average, show the opposite.
However, there are many exceptions to this picture \citep[see][for more discussion]{2024MNRAS.531.1056R}.

Another clear feature seen in Fig.~\ref{fig:offset-filters-set-teq_prd} is the existence of very large phase-curve offsets 
seen in filters \hstwfc, \nircam\ F090W, F115W, F150W, F162M, F164N, F210M, and F212N for, on the average, low gravity and low $\teq$ models.
These offsets can reach values as high as $\Delta\phi$$\approx$120$\degr$. This is partly because these models lack
\tio/\vo\ opacity and, as a result, most of the light at these wavelengths originate from deep layers (p$\approx$1~bar) both on day and night sides. 
Additionally, these wavelengths lack significant contribution from the reflected light, too.
We emphasize that the phase curves in filters that show such large offsets are the ones that have very small flux variability
across phases that cover eastern hemisphere, similar to the \hstwfc\ curve in Fig.~\ref{fig:lc-cf-example}. 
Thus, in real observations the data reduction steps and noise could bias results toward smaller or higher offset values in such cases.
The large scatter seen in such filters such as \hstwfc\ and \nircam\ passbands between 1~\mum\ and 2~\mum\ are likely because they specifically cover 
strong temperature sensitive \hho\ bands. For models without \tio/\vo, the radiation in these filters originates from deep layers
where the flow is affected by gravity and rotation, for example, so that temperature distribution varies noticeably between the models and this is
best manifested in temperature sensitive water bands.
The rest of filters (\spitzer, \miri, \nirspec, and \nircam\ filters with wavelengths $\lambda$$>$3~\mum) demonstrate very similar range of offset values 
($\Delta\phi$$<$60$\degr$). The marginal retrograde (i.e., negative) offsets are found only in short wavelength filters, i.e., those that are sensitive
to the stellar reflected light.

The predicted contribution functions are summarized in Figs.~\ref{fig:cf-tess-cheops-hst-spitzer}, \ref{fig:cf-nirspec-miri}, \ref{fig:cf-nircam1}, and ~\ref{fig:cf-nircam2}, respectively.
They demonstrate that the effective light formation depth shifts toward deeper layers as planet gravity increases for all filters considered. 
On the dayside, filters that best probe high altitudes for the models with $\teq$ around 1600~K
are \tess, \cheops, \hstgfourthreezero, most of \spitzer, \nirspec, \miri, and \nircam\ filters, 
while \nircam\ filters F460M, F466N, F470N, and F480M can probe very high altitudes for even hotter planets. 
The deep atmospheric layers on the dayside can be best probed with such filters such as \tess, \cheops, \hst\ filters, 
\nirspec\ Prism, and \nircam\ filters F070W, F090W, and F115W, respectively. 
On the nightside, high altitudes are best probed by, for example, \hstgfourthreezero, \spitzeriracseven, \miri\ F770W,
\nircam\ F070W, F250M, and F277W, respectively, while the best sensitivity to low altitudes is observed in the following filters:
\tess, \cheops, \hstwfc, \nirspec\ Prism, \nircam\ F115W, F162M, and F164N.

The phase-curve amplitudes calculated from GCMs are plotted in Fig.~\ref{fig:af}.
Similar to previous studies, we write the phase-curve amplitudes as $A_{\rm f}$=($F_{\rm max}$ $-$ $F_{\rm min}$)/$F_{\rm max}$. 
Defined this way, the amplitude can reflect the heat redistribution efficiency between day and night sides of a tidally locked planet
at atmospheric depths probed by a photometric filter: it varies between 0 (full redistribution) and 1 (no redistribution).
Figure~\ref{fig:af} shows that, in all filters, the phase-curve amplitude increases with $\teq$ thus indicating that strongly
irradiated models have inefficient heat redistribution which is also manifested
as a decreases in the phase-curve offsets described above. Large amplitudes can be seen in short wavelength filters
for models with $\teq$$\geqslant$1400~K, but especially in filter \hstgfourthreezero\ for which $A_{\rm f}$$\approx$1 for all models.
The reason for this is that \hstgfourthreezero\ only probes wavelengths shorter than 0.6~\mum. There, the planet emission on the night
side is very weak, while filters such as \cheops\, which also cover wavelengths $\lambda$<0.6~\mum\, additionally collect light 
from longer wavelengths where the nightside flux is much stronger (see Fig.~\ref{fig:t-flux-filters}) which decreases the value
of the phase-curve amplitude.
Within many filters there is also a large scatter of amplitudes with $\teq$ which can be best seen in such filters such as
\spitzeriracthree, \nircam~F162M, F164N, F335M~--~F410M.
This scatter arises from variations in rotation period and surface gravity among individual planets.
Another interesting feature is the global decline of the phase-curve amplitude
between short and long wavelengths best seen on Fig.~\ref{fig:af} as one goes from \cheops\ up to \spitzermipstwentyfour.
This happens because the planet emission fades toward long wavelengths on the dayside, and the flux contrast between hot (day) and cold (night) sides
decreases \citep[see also, e.g.,][]{2021MNRAS.501...78P}. 
A similar trend can be seen between \nirspec\ and \miri~F1140C filters, and to a lesser extent between \nircam\ filters F070W and F164N, respectively.
Finally, there is no strong dependence on gravity. However, massive planets, on the average, tend to have larger amplitude,
in agreement with \citet{2024MNRAS.531.1056R}.

\subsection{Comparison with observations}

\begin{figure*}
\centerline{
\includegraphics[width=\hsize]{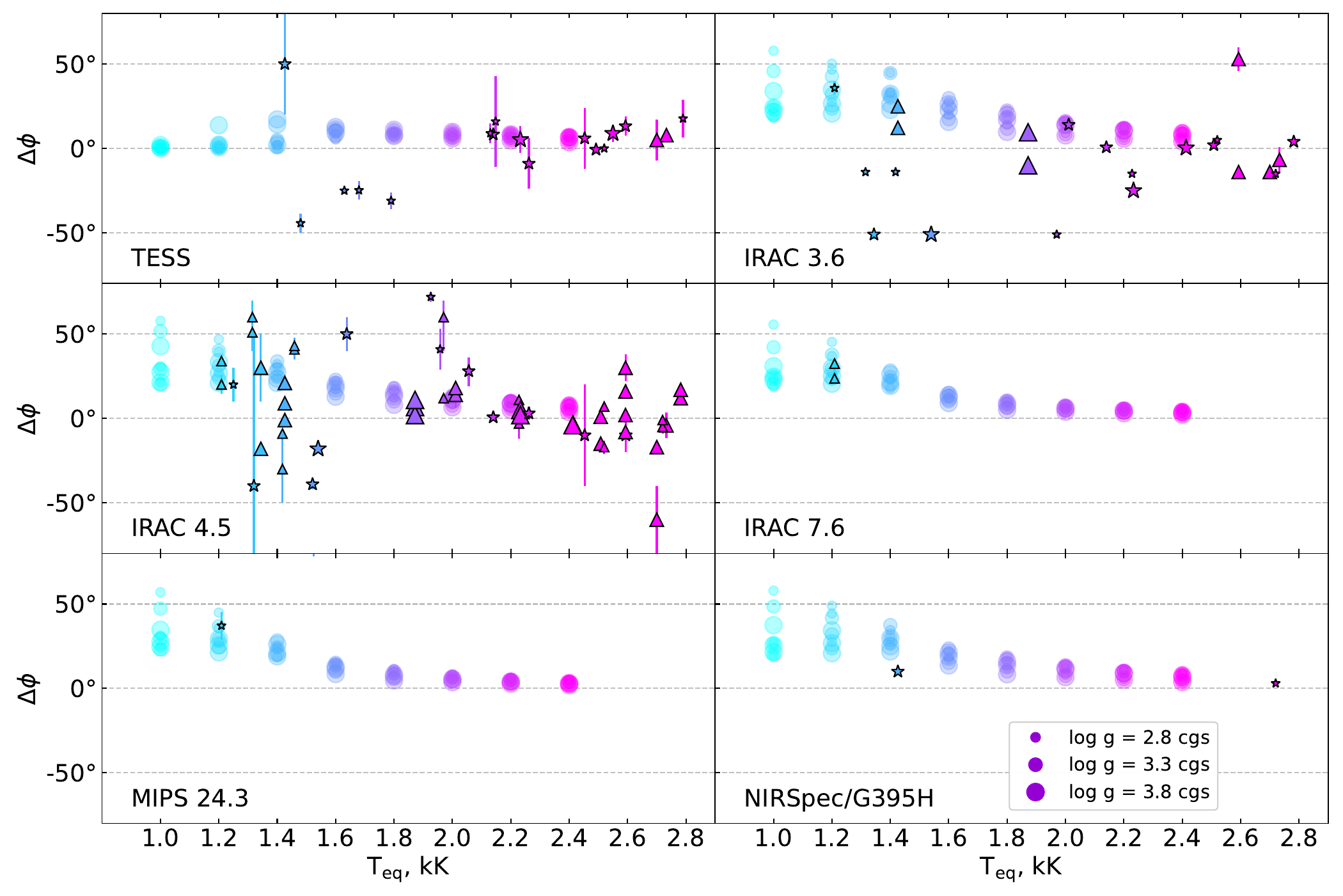}
}
\caption{Comparison between predicted and observed phase-curve offsets in selected filters.
Models are shown with circles and the size of the symbols corresponds to the surface gravity.
The observed data points are marked with star symbols for planets with a single measurement, and with triangles otherwise.
In the latter case each triangle corresponds to an individual estimate, respectively.
The color coding is according to equilibrium temperature.
Error bars for the observed data points were taken from original literature sources when available.}
\label{fig:offset-obs}
\end{figure*}

We now compare model predictions with selected observations of exoplanet phase curves,
which we indicate in Fig.~\ref{fig:offset-obs} with star symbols that are color-coded according to 
the planet's surface gravity. The observed offsets were taken from different literature sources:
\citet{2025ApJ...982..159S,2025AJ....169...32D,2025ApJ...983L..13L,2024A&A...687A.144K,2024ApJ...969L..32C,
2022A&A...668A..17S,2022PSJ.....3..255E,
2021AJ....162..127W,2021AJ....162..158M,2021AJ....162..218C,2021AJ....162...62D,2018haex.bookE.116P,
2016Natur.532..207D,2015ApJ...804..150E,2016ApJ...823..122W,2009ApJ...690..822K,2012ApJ...754...22K,2012ApJ...755....9S,
2014ApJ...790...53Z,2010ApJ...723.1436C,2012ApJ...747...82C,2015ApJ...811..122W,2018ApJ...855L..30A,2013MNRAS.428.2645M,
2014Sci...346..838S,2017AJ....153...68S,2015ApJ...802...51H}. Thus multiple measurements from different works
could be presented in Fig.~\ref{fig:offset-obs} for a single planet (e.g., filters \spitzeriracthree\ and \spitzeriracfour).
Keeping in mind that the observed data is very sparse in some filters, nevertheless, we do not observe strong trend of increasing phase-curve offset
as $\teq$ decreases. However, if one ignores negative offsets
in filters \tess, \spitzeriracthree, and \spitzeriracfour\ for which many independent measurements exist, the models tend to agree with the data.
Still and contrary to the model predictions, there are many instances where data show large negative
offsets. In conclusion, comparing models with observations remains challenging, 
both because of the scarcity of phase-curve offset measurements and the significant discrepancies between individual measurements
(see Fig.~\ref{fig:offset-obs-irac4.4} for the example of a scatter in individual measurements for the \spitzeriracfour\ filter).
More work is clearly needed to determine whether these discrepancies arise 
from data-analysis issues or reflect real physical variability.

\subsection{Resolving the vertical structure of atmospheric flows with high-resolution spectroscopy}

High-resolution spectroscopy is a rapidly improving technique to study the chemical and physical
structure of planetary atmospheres. Using three models as a proxy of a hot Jupiter 
($\teq$$=$1000~K, $\teq$$=$1600~K, and $\teq$$=$2200~K with $\mplanet$$=$1.36$\Mjup$ orbiting $\mstar$$=$1.1$\Msun$ star),
we simulated spectra in the H (1.5~\mum~--~1.7~\mum) and K (2.3~\mum~--~2.5~\mum)
bands in order to test high resolution spectroscopy for separating flows at
different atmospheric depths using emission observations. We do it by disk integrating spectra 
of the planet at phases $\phi$$=$0.5 (dayside) and $\phi$$=$0.0 (nightside) including molecular opacity
only due to most abundant molecules \hho, \chhhh, \coo, \co, \hcn, and \nhhh\ in order to save computation time. 
The flow velocity were taken from the original SPARC/MITgcm models and vector transformed to the observer rest frame.
We do not prescribe any limb-darkening coefficient because we directly calculate specific intensities for a set of view angles and then interpolate
between them to obtain a value for each surface pixel. The Doppler shift due to rotation is also taken into account for each pixel at this step.
We then use a template spectrum for each of the molecule that we try to detect, which we choose to be \hho, \chhhh, and \co, 
and cross correlate these templates against the simulated observations.
Note that we do not attempt to reproduce any of data reductions steps and do no attempt to simulate relevant noise sources in our synthetic observations.
Instead, our goal is to quantify the displacement of line profiles using cross-correlation function (CCF), and compare results 
with the capabilities of modern spectroscopic instruments. 
The template spectra were computed using the average T(p) profile for the day and night sides, respectively. 
Each template includes line opacity due to the molecule under investigation, plus all relevant continuum opacity sources mentioned in Sec.~2.
Before calculating CCF, we convolve all spectra to have same spectral resolution of R$=$$100\,000$, which is the typical
resolving power of the \criresplus\ instrument (The CRyogenic InfraRed Echelle Spectrograph Upgrade Project)\footnote{https://www.eso.org/sci/facilities/develop/instruments/crires\_up.html}, for example.
We then normalize all spectra using low order polynomial. We also apply corresponding rotation broadening using a linear limb darkening law
with coefficient $\epsilon$$=$1 \citep[see, e.g.,][]{2023A&A...672A.107Y}, obviously to the template spectra only.
The rotation broadening for this test case increases with planet $\teq$ because the distance to the star decreases and hence orbital periods decrease, too,
and amounts to $\upsilon_{\rm eq}$$=$0.65~\kms\ ($\teq$$=$1000~K), $\upsilon_{\rm eq}$$=$2.2~\kms\ ($\teq$$=$1600~K),
and $\upsilon_{\rm eq}$$=$7~\kms\ ($\teq$$=$2200~K), respectively.
Note that the accuracy of the CCF position in the absence of noise is defined by the assumed step size of the velocity shift, which we chose to be 0.1~\kms.

\begin{figure}
\centerline{
\includegraphics[width=\hsize]{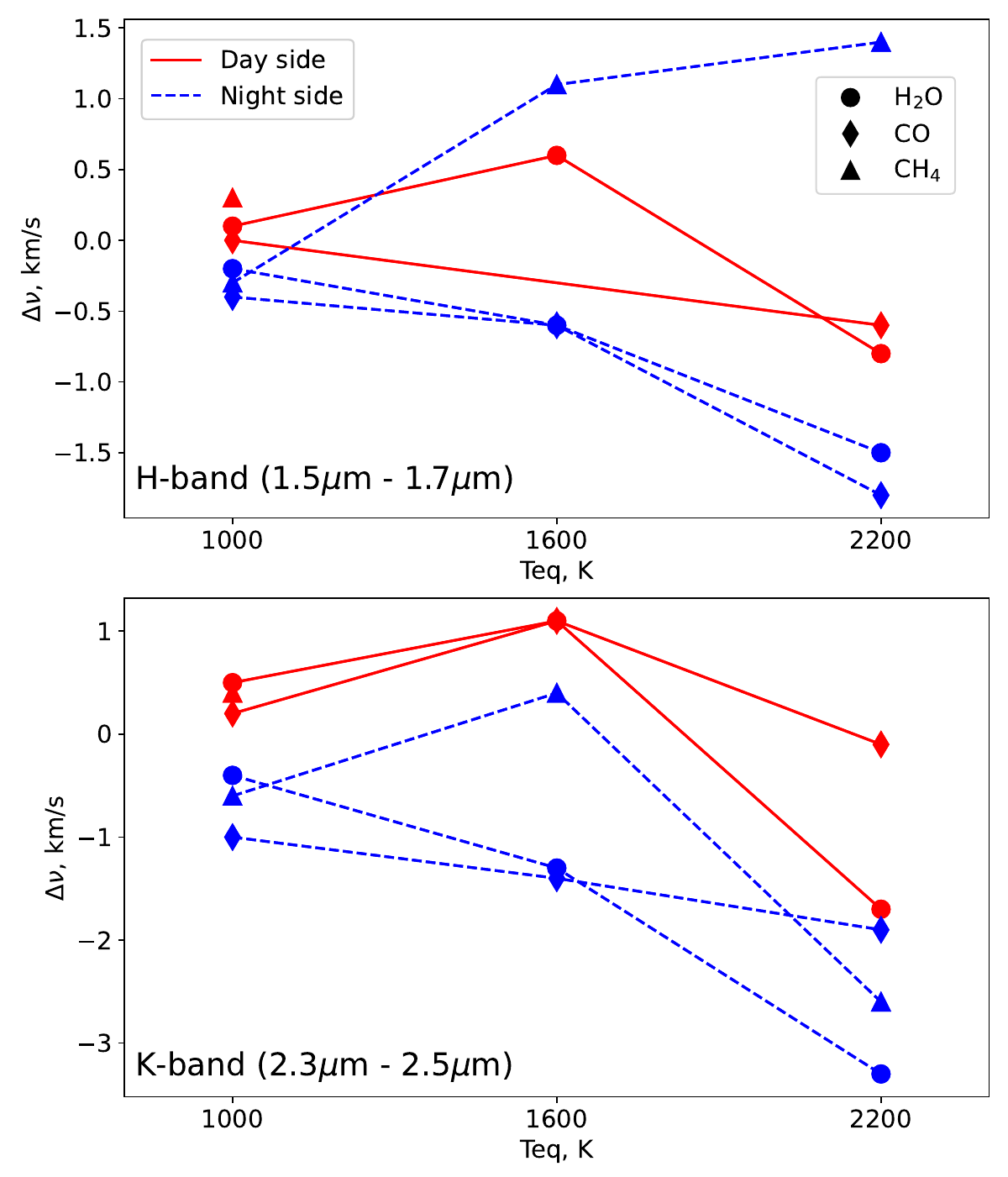}
}
\caption{Radial velocity shifts of selected molecules estimated from the three models with $\teq$$=$1000~K, $\teq$$=$1600~K, and $\teq$$=$2200~K,
in the H band (top panel) and the K band (bottom panel), respectively. For all models, $\mplanet$$=$1.36$\Mjup$ and $\mstar$$=$1.1$\Msun$ were assumed.
Solid red and dashed blues lines connect measurements on the day and night sides, respectively. Note that in some cases the signal could be detected
and hence the respected data points are not shown.
}
\label{fig:drv_mols}
\end{figure}

Our calculations show that, while local flow velocity can be comparable to the rotation velocity, the net effect on the CCF is indeed small, 
but can be non-negligible in some cases, as illustrated in Fig.\ref{fig:drv_mols}.
For example, on the dayside of the $\teq$$=$2200~K model, a noticeable relative velocity shift of $\delta\upsilon$$=$$-$0.9~\kms\ is found between
\hho\ lines located in the K and H bands, and an even more stronger shift of $\delta\upsilon$$=$$-$1.6~\kms\ in the K band between \hho\ and \co\ lines, respectively.
For the $\teq$$=$1600~K model, only a $\delta\upsilon$$=$$+$0.5~\kms\ between \hho\ lines located in the K and H bands is detected.
No significant displacement could be measured on the dayside of the $\teq$$=$$1000$~K model.

On the nightside and for the $\teq$$=$2200~K model, we find a relative shift of $\delta\upsilon$$=$$-$1.8~\kms\ between \hho\ lines in the K and H bands,
and $\delta\upsilon$$=$$-$1.4~\kms\ between \hho\ and \co\ lines in the K band, respectively. In addition, the lines of \chhhh\ show strong
relative displacement in the K and H bands of $\delta\upsilon$$=$$-$4~\kms\ due to a velocity sign change (negative in the K band and positive
in the H band, respectively). For the colder $\teq$$=$1600~K planet the lines of \hho\ and \co\ in the K and H bands have very similar
relative displacement of $\delta\upsilon$$\approx$$-$0.7~\kms. It is the same for the \chhhh, too, but the shift is positive.
On the nightside of the $\teq$$=$1000~K model, only \co\ lines show a weak shift of $\delta\upsilon$$=$$-$0.6~\kms\ between K and H bands,
and similar shift between \hho\ and \co\ lines in the K band, respectively.

The typical velocity resolution in the K band of the \criresplus\ is $\delta\upsilon$$\approx$3~\kms, and current CCF techniques already can reach
the uncertainty of $\delta\upsilon$$\approx$1~\kms\ in measuring positions of molecular lines \citep[see, e.g.,][]{2025A&A...693A..72L,2024A&A...688A.206C,2023A&A...672A.107Y}.
Therefore, the use of different spectral regions appears promising in resolving vertical structure of atmospheric winds in hot Jupiters
from high-resolution spectroscopy. In addition, combining emission and transmission observations can help to improve the accuracy in measuring atmospheric winds
at high altitudes. It also naturally separate flows at dusk and dawn hemispheres that facilitates the measurements of line position. This can be used
in resolving vertical structure of winds, as was recently shown by \citet{2025Natur.639..902S,2024ApJ...975....9K}.
Finally note that detecting spectroscopic signals from high-resolution spectroscopy on the nightside of HJs remains challenging 
because of the weak line strength that originates in the approximately isothermal atmospheric regions \citep[see, e.g.,][]{2025A&A...698A..31C,2025ApJ...986...63W}.

\begin{figure}
\centerline{
\includegraphics[width=\hsize]{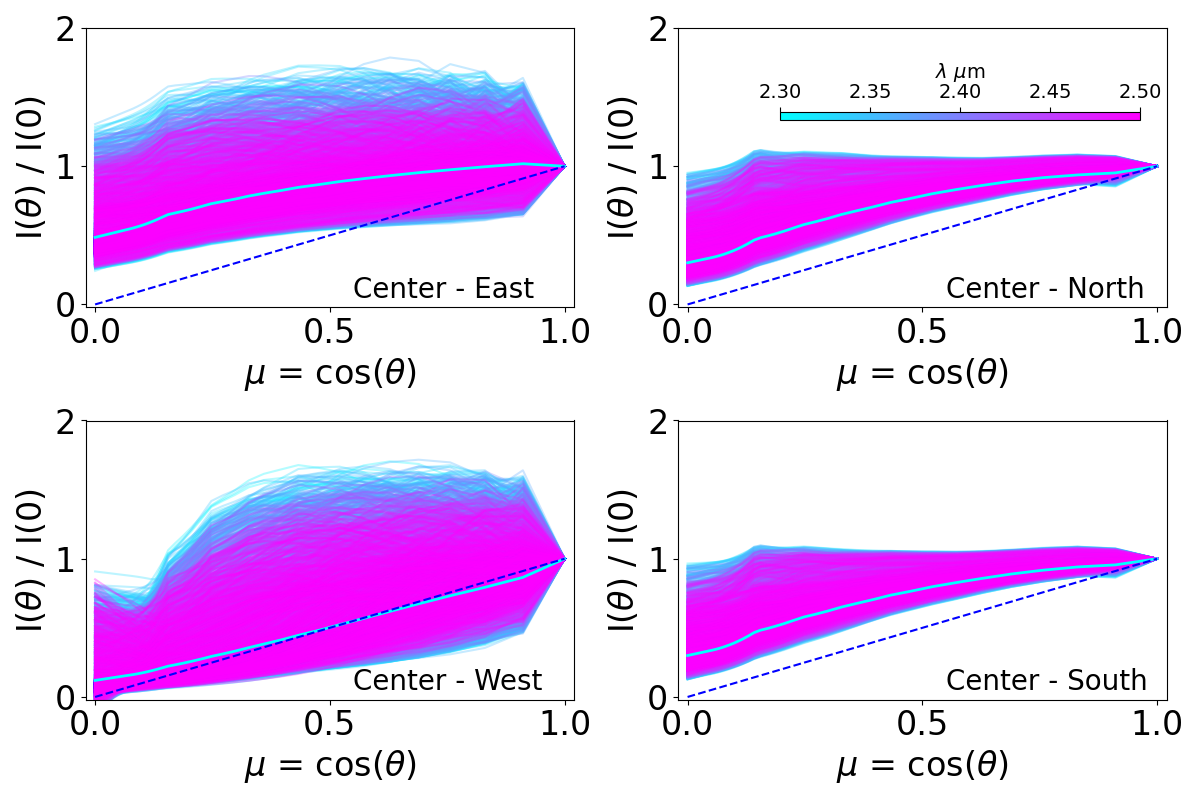}
}
\caption{Predicted dayside limb darkening curves calculated from  $\teq$$=$2200~K, $\mplanet$$=$1.36$\Mjup$, $\mstar$$=$1.1$\Msun$ model
in the K band (2.3~\mum~--~2.5~\mum) with resolution $R$$=$100\,000.
Each plot corresponds to one of the four selected directions along which the intensity of radiation
was calculated between center ($\mu$$=$1) and limb ($\mu$$=$0)
in the center~--~east (top left), center~--~north (top right), center~--~west (bottom left), 
and center~--~south (bottom right) directions, respectively. The thick solid line is the wavelength
average, and the dashed line is the linear limb darkening law with the coefficient $\epsilon$$=$1.
The color coding is according to wavelength. 
}
\label{fig:limb-darkening-example-k-band}
\end{figure}

\begin{figure}
\centerline{\tiny \hspace{0.45cm} H-band (1.5~\mum~--~1.7~\mum) \hspace{1.5cm} K-band (2.3~\mum~--~2.5~\mum)}
\centerline{
\includegraphics[width=0.5\hsize]{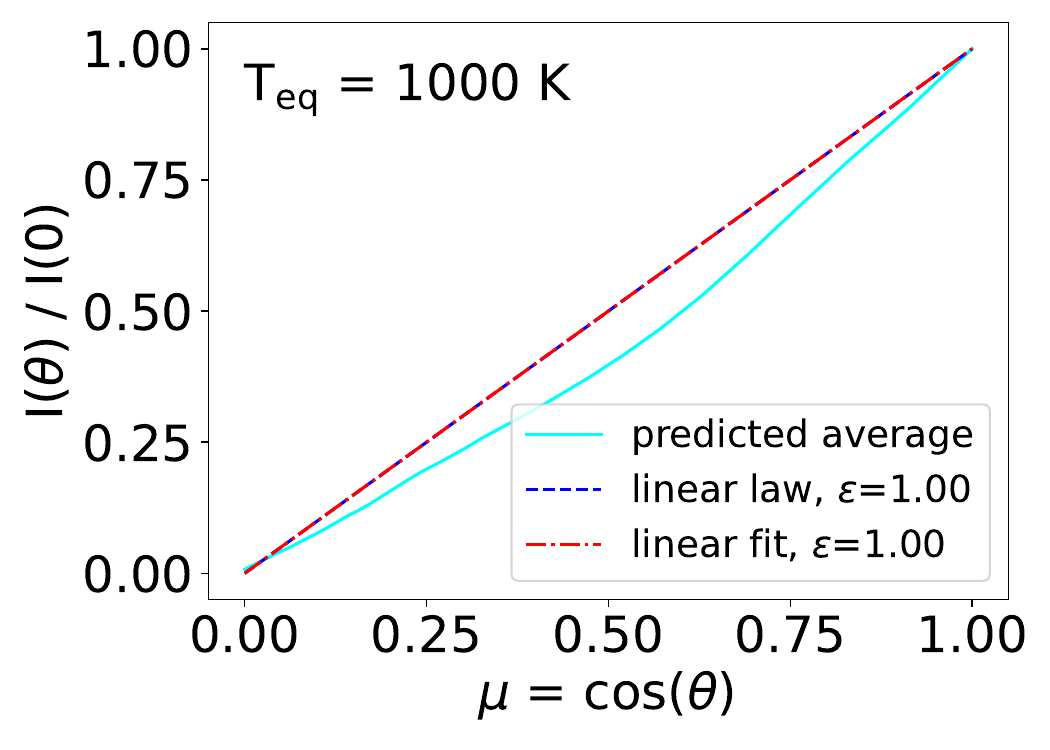}
\includegraphics[width=0.5\hsize]{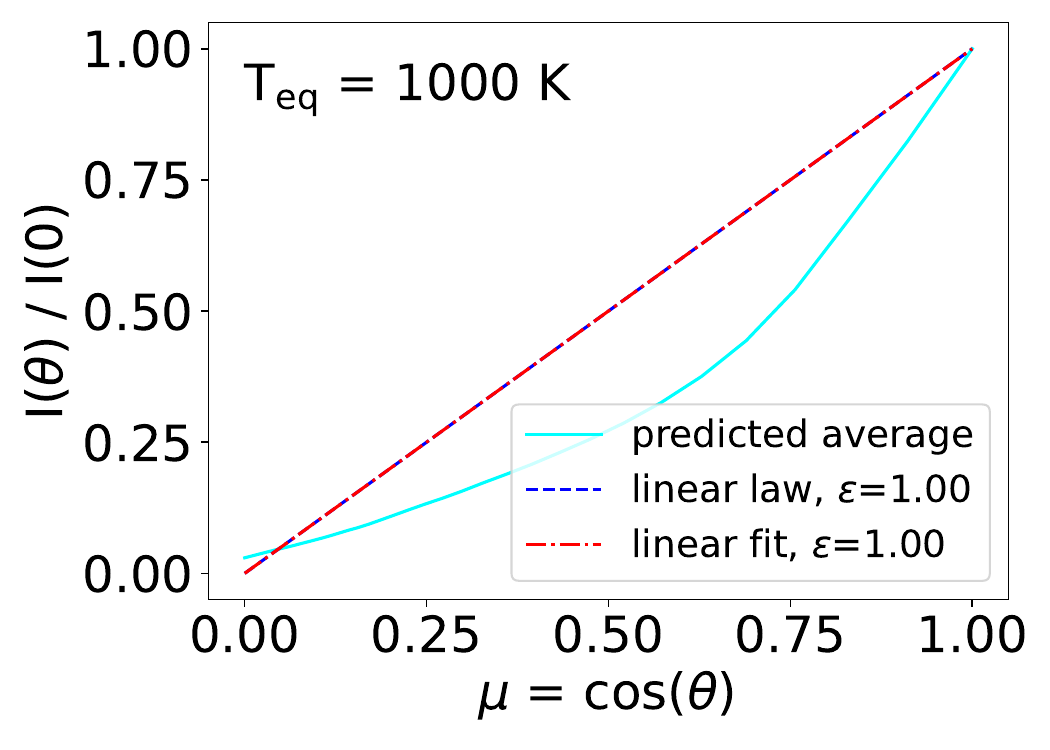}
}
\centerline{
\includegraphics[width=0.5\hsize]{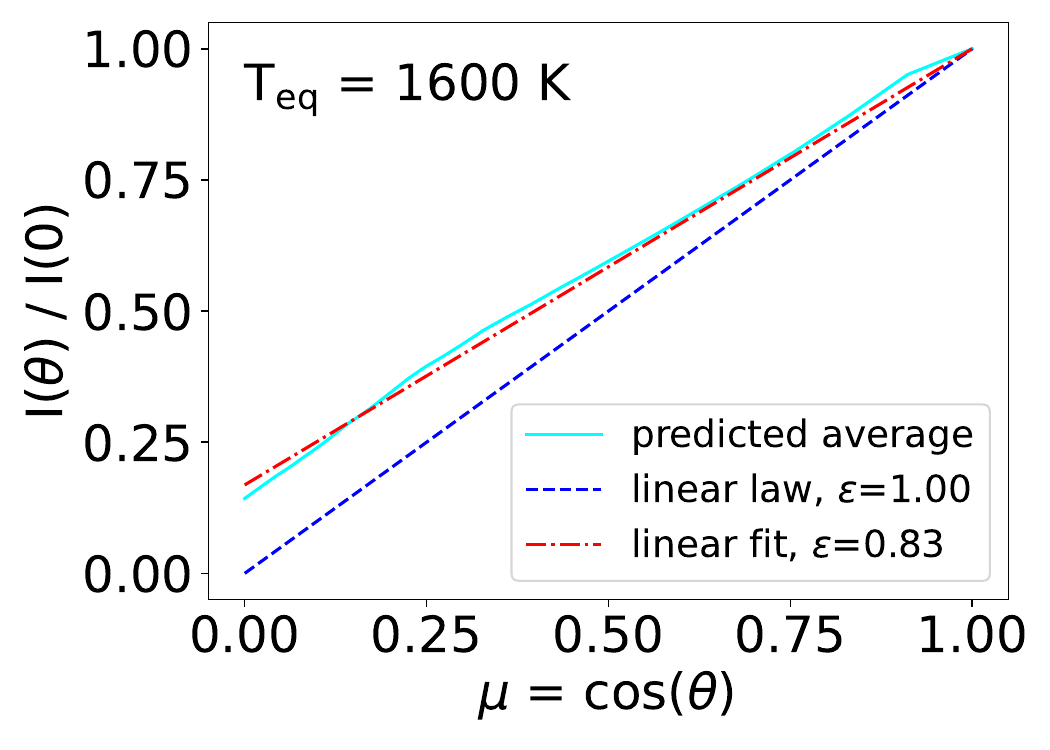}
\includegraphics[width=0.5\hsize]{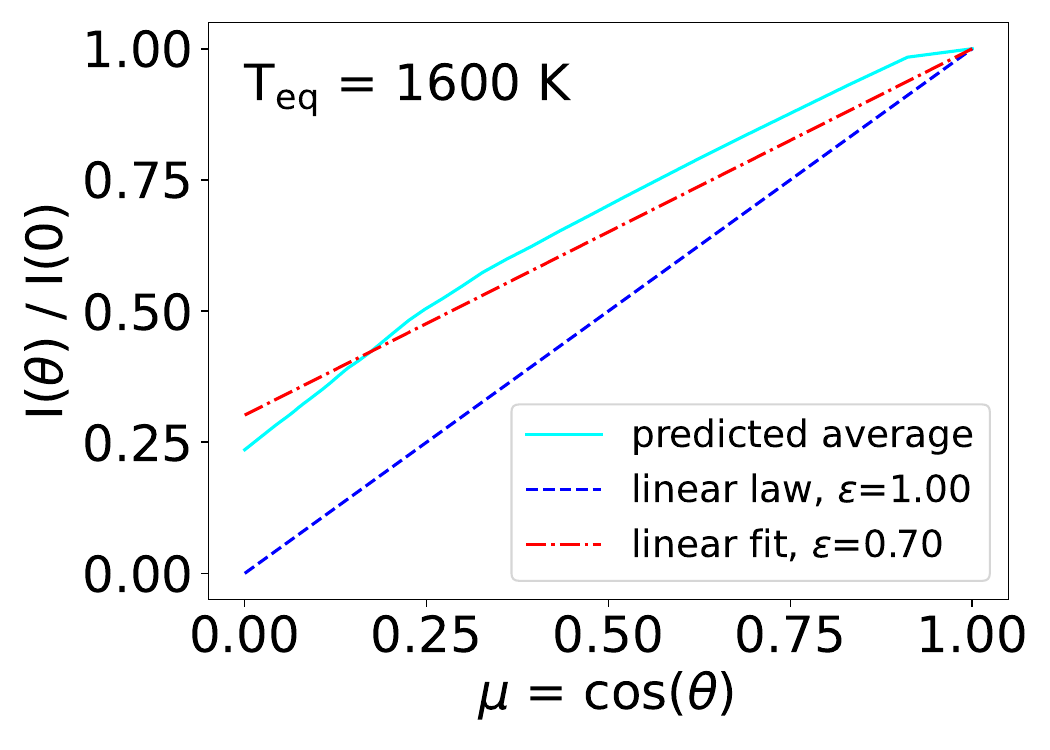}
}
\centerline{
\includegraphics[width=0.5\hsize]{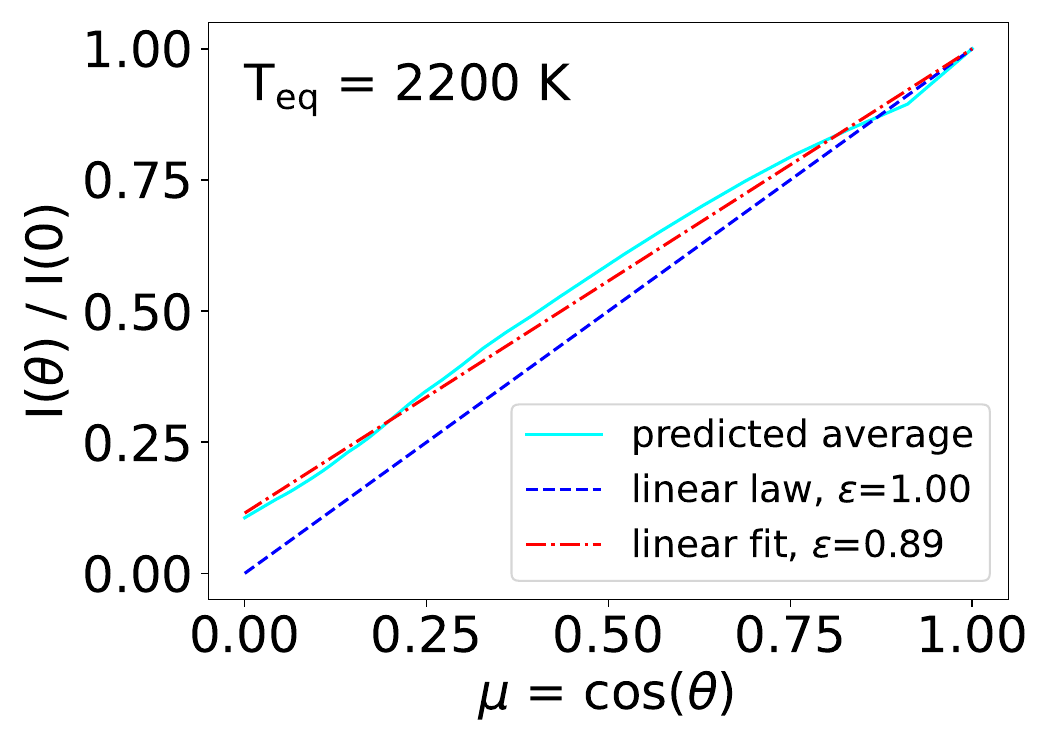}
\includegraphics[width=0.5\hsize]{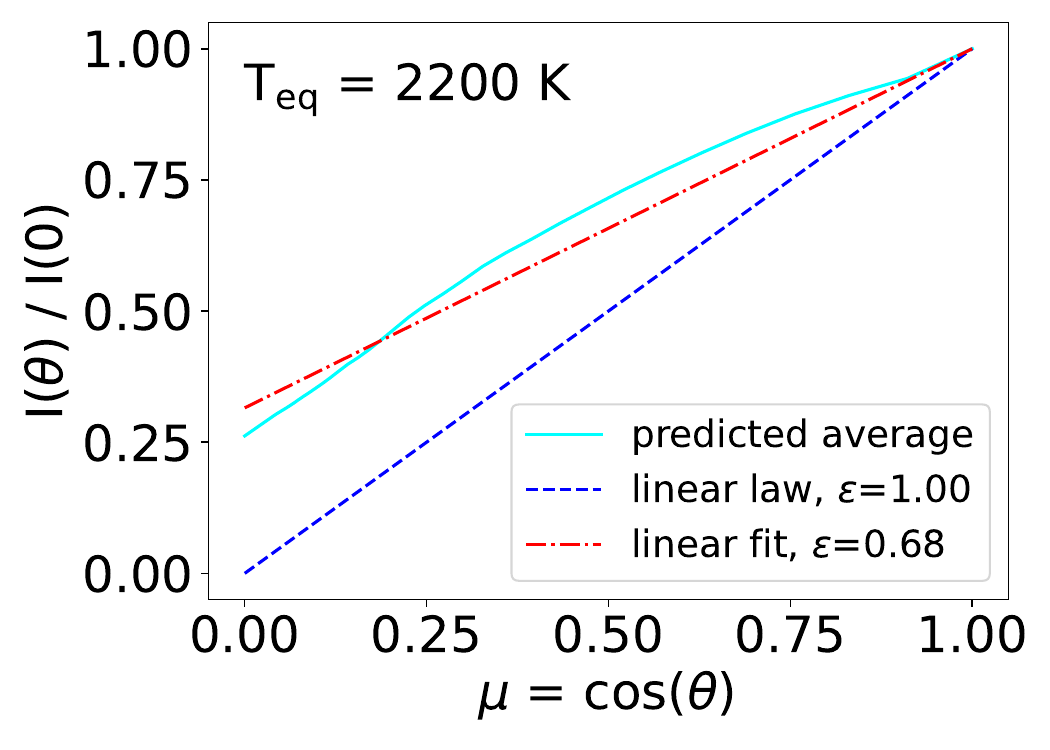}
}
\caption{Predicted limb darkening curves calculated from models with (from top to bottom) $\teq$$=$1000~K, $\teq$$=$1600~K,
and $\teq$$=$2200~K, respectively. For all models, $\mplanet$$=$1.36$\Mjup$ and $\mstar$$=$1.1$\Msun$ was assumed.
The two columns present plots of limb darkening curves averaged over all wavelengths 
and twenty four different directions between the center and limb of the visible planet disk
in the H (1.5~\mum~--~1.7~\mum, left column) and K (2.3~\mum~--~2.5~\mum, right column) bands, respectively.
The linear limb darkening law with coefficient $\epsilon$$=$1 is shown as a dashed line. 
The dash-dotted line shows the linear fit to the predicted curve (solid thick line) 
and the values of the derived best-fit coefficient are listed in the legends to the plots.
}
\label{fig:limb-darkening-average}
\end{figure}

Lastly, a nonsymmetric brightness of the planet surface can result in rather strong variations of specific intensity
between center and the limb. However, for the simplicity, most, if not all, current works that utilize
CCF for the modeling of atmospheric signals assume that the dayside radiation is dominated by a localized hotspot only, 
while limb regions are colder with little contribution to the total light. 
Thus, to calculate rotation broadening for the template spectra,
a linear limb darkening law has frequently been applied \citep{2011A&A...531A.143D}
with a coefficient $\epsilon$$=$1 to suppress the contribution from the limb \citep{2023A&A...672A.107Y}.
To test whether this assumption is appropriate we used models for the planets with $\mplanet$$=$1.36$\Mjup$, $\mstar$$=$1.1$\Msun$,
and three equilibrium temperatures $\teq$$=$1000~K, $\teq$$=$1600~K, and $\teq$$=$2200~K and calculated specific intensity 
between center and the limb in different directions and recovered the limb darkening law in the H and K bands.
Our calculations show that specific intensity can deviate strongly from the linear law at individual wavelengths, which is best seen
in the K band thanks to the presence of many strong \co\ lines. An example of this is illustrated in 
Fig.~\ref{fig:limb-darkening-example-k-band} for the $\teq$$=$2200~K model.
Specifically, the intensity toward the limb can become stronger than that at the center of the disc in some wavelengths and in 
the center~--~east (dusk) and center~--~west (dawn) directions. In the latter case and for angles $\mu$$<$0.1 
a clear limb brightening can be seen.
This is due to the redistribution of the hot atmospheric material along the equator
and the presence of the temperature inversion layer that increase the intensity of the molecular lines seen in emission at high altitudes.
The intensity originating from the directions toward the poles is never stronger than that of the disk center because poles remain relatively cold
and lack strong temperature gradients.
Doing the same excises for different models and using now twenty four different limb points for a more accurate representation
of the average limb darkening curve we recovered plots presented in Fig.~\ref{fig:limb-darkening-average}.
The average values appear to coincide closer to the linear law in the H band (note that no good linear fit could be
obtained for the $\teq$$=$1000~K model in both H and K bands, respectively, so that the best fit linear coefficient is tight to $\epsilon$$=$1).
However, the linear law seem to underestimate the contribution from most of the planet surface
in the K band in all three models thanks to the presence of many strong \co\ lines seen in emission. 
We caution that these results are based upon only three case studies
and they cannot be easily extrapolated neither to other models from our grid nor to the real planets. 
By doing this we only show that, at least at some spectral windows, 
the assumption $\epsilon$$=$1 may not be accurate. For the practical purpose, the choice of the coefficient $\epsilon$
was reported to have a relatively small impact on the measurements of atmospheric winds, amounting to $\approx$1-2~\kms\ accuracy 
in the K band for the WASP-43~b \citep{2023A&A...678A..23L}. However, it may become important when measuring a velocity difference
between various spectral bands in an attempt to resolve flows with altitude.

\section{Discussion}

\subsection{Modeling approach for planetary emission}

In this work, we utilized the same grid of GCMs that was used in \citet{2024MNRAS.531.1056R}.
The grid artificially excludes opacity due to \tio/\vo\ for all models with $\teq$$\leqslant$1200~K
because, as it is argued in \citet{2024MNRAS.531.1056R}, there is clear evidence of the absence of any
significant \tio/\vo\ absorption from the lack of thermal inversions in the atmospheres of HJs around those temperatures.

\begin{figure}
\centerline{
\includegraphics[width=\hsize]{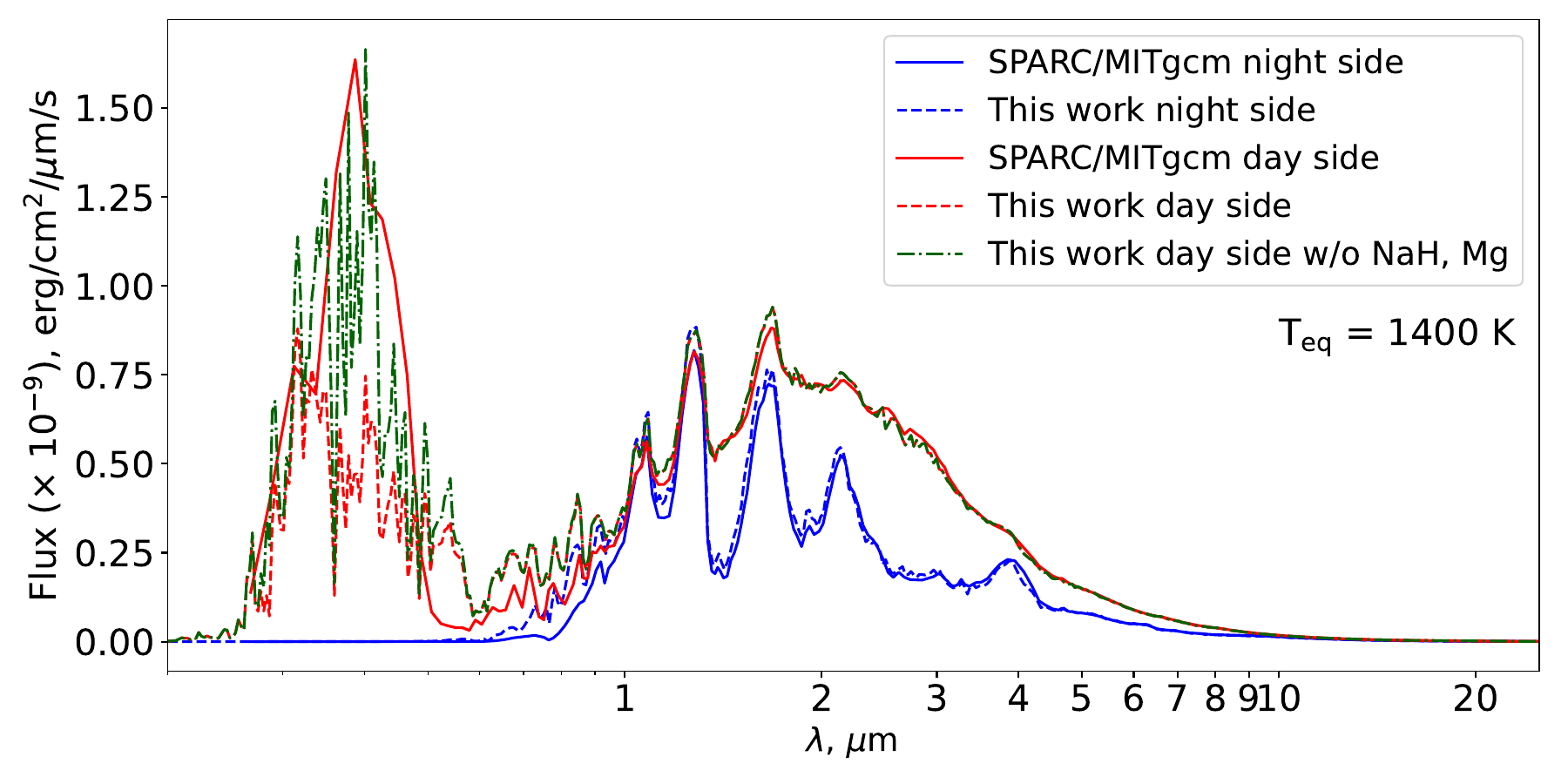}
}
\caption{Comparison between theoretical emergent flux predicted by an original GCM model from the grid
by \citet{2024MNRAS.531.1056R} and this study. The model parameters are $\teq$$=$1400~K, $\mstar$$=$1.1~$\Msun$, $\logg$$=$3.3~dex.
We show the disk integrated flux at orbital
phases $\phi$$=$0.0 (nightside) and $\phi$$=$0.5 (dayside), respectively. See the figure legend for details.
}
\label{fig:flux-teq1400_nah_mg}
\end{figure}

In the calculations of the planetary emission we utilized opacity due to 86 atomic and molecular species 
which is more complete compared to the set of opacity used in SPARC/MITgcm. This is likely not 
the issue for predicting the accurate temperature structure of the atmosphere which otherwise only depends upon
the balance between frequency integrated absorption and emission coefficients and thus requires inclusion 
of the opacity due to main absorbing species. But minor opacity source that do not play a role in global energy balance
could still impact the flux in photometric filters that cover localized wavelength regions.
For instance, opacity due to lines of \ion{Mg}{i} and NaH is important in the short wavelength region where it dims 
the emission due to the stellar reflected light in models with low $\teq$. 
{As an example, Fig.\ref{fig:flux-teq1400_nah_mg} shows a comparison of the emergent flux
for the $\teq$$=$1400~K model ($\mstar$$=$1.1~$\Msun$, $\logg$$=$3.3~dex) from the grid by 
\citet{2024MNRAS.531.1056R} and this study
(see Fig.~\ref{fig:flux-teq} for more comparison between models with different $\teq$).
We also show an additional calculation where we excluded NaH and \ion{Mg}{i} opacity to highlight its role
at short wavelengths. Without this opacity our predicted flux on the dayside is closer to the one
from the grid by \citet{2024MNRAS.531.1056R}. On the nightside our calculation predict the higher flux
compared to the flux generated by GCM in the wavelengths between 0.6~\mum\ and 1~\mum. 
This is most likely due to a different approach to how the condensation of \tio\ and \vo\ were 
treated when computing opacity for the GCMs and the present study where we used the latest version of the \fastchem\ code.
A choice of the \tio/\vo\ line lists also contributes to the observed difference. In our study we made use of the latest \textsc{ToTo}
and \textsc{VOMYT} line lists for \tio\ and \vo\  by \citet{2019MNRAS.488.2836M} and \citet{2016MNRAS.463..771M}, 
while GCMs utilized the line lists by \citet{1998FaDi..109..321S} and \citet{1998A&A...330.1109A}, respectively. 
The latter line lists result in higher opacity in the optical compared to the former \citep[see, e.g.,][]{2026ApJ...997..365W}.
In the infrared ($\lambda$>1~\mum) the predicted flux agree well.
The total (i.e., integrated) flux on the day and night sides 
differ by $\approx$3\%~--~5\%\ between the two calculations (with the largest difference found for $\teq$$=$1600~K model).
Thus, due to a rather small impact on the total radiative energy balance, 
we do not expect these differences to significantly modify the atmospheric flows predicted by GCMs.
Besides, there are other processes that were not included in our GCMs (e.g., photo-dissociation, clouds, etc.) 
that may have stronger impact on the atmospheric structure compared to these secondary opacity sources.}

\subsection{Predicted phase curves}

Overall, the phase curves calculated in this study are in good agreement with the analysis by \citet{2024MNRAS.531.1056R}, 
but we aimed to analyze flux in individual filters rather than looking at bolometric values. 
We find that not all filters show behavior which is similar to the bolometric flux.
Additional parameters~--~rotation and gravity~--~also play a role.
For instance, in models with short rotation periods (i.e., $\mstar$$=$0.8$\Msun$), most of the filters
show a steady decrease of the offset as $\teq$ rises (see Fig.~\ref{fig:offset-teq-logg-ms0.8}).
Only short wavelength filters demonstrate pattern similar to the bolometric flux: the offset
first increases when $\teq$$\leqslant$1400~K, and then decreases. 
In planets with longer rotation periods the increase of the offset for the $\teq$$\leqslant$1400~K
could be seen in more filters and mostly in low gravity planets, while planets with highest mass
can still show a steady decrease of the offset with $\teq$ in some cases 
(e.g., some \spitzer\ and \miri\ filters, see Fig.~\ref{fig:offset-teq-logg-ms1.1} and Fig.~\ref{fig:offset-teq-logg-ms1.5}).
Hence we notice a rather complex behavior of the offsets in individual filters compared to the bolometric flux.

The amplitude of the phase curve has a complex behavior in some filters as well. Similar to \citet{2024MNRAS.531.1056R}, 
we notice a general trend of decreasing the amplitude as temperature decreases.
But this is not always the case for some filters. As was noted above, filter \hstgfourthreezero\ does not show any noticeable
dependence on any planet or stellar parameter, and the \cheops, and \nircam\ F070W filters can have similar amplitude 
in some models with $\teq$$\leqslant$1400~K. Additionally, \nircam\ filters F335M, F356W, and F360M show similar high amplitude
for both hot and cold models as planet gravity increases.
Finally notice a clear trend of the amplitude being the largest in short wavelength filters and decreases as wavelength increases
for relatively hot planets with $\teq$$\geqslant$1600~K as a result of the decrease in the flux contrast between day and night sides 
toward long wavelengths. For the cooler planets, the reflected light from clouds could become important 
and increase the day/night contrast even further.

\subsection{Phase-curve offset: Predictions versus observations}

The results of our study demonstrate a clear correlation between the planet temperature
and offset value: as $\teq$ increases, the offset decreases.
The gravity of the planet only has a secondary effect, except for some
filters such as \nircam\ F090W, F140M, F162M, F164N, F150W, F210M, F212N, \hstwfc, and 
for the short orbit low temperatures planets additional filters \nircam\ F323N, 335M, \miri\ F770W, and \spitzeriracseven, respectively.
This agrees with previous studies \citep[e.g.,][]{2002A&A...385..166S,2024MNRAS.531.1056R} and their conclusion that in low temperature HJs
the atmospheric flows are efficient in moving hot gas away from the substellar point in deep layers where the advection
timescale is shorter than the radiative one.

Atmospheric metallicity is expected to play an important role in shaping phase-curve properties 
through its influence on opacity and hence radiative timescales. 
In this work we did not investigate the effect of metallicity to keep computations feasible, and considered only solar metallicity models. 
This is further justified by recalling that our primary goal is to test the sensitivity 
of individual photometric filters to the atmospheric flows. Metal rich models would show smaller phase-curve offsets
thus limiting the range of altitudes probed by photometric filters, as discussed in \citet{2024MNRAS.531.1056R} (see their Fig.~8).
However, when looking at individual filters,
the decrease in the offset value is only observed for models with $\teq$$\geqslant$1600~K. 
For colder planets, the offset decreases with increasing metallicity only at long wavelength filters. 
At short wavelengths, on the other hand, the offset value can in fact increase in metal rich planets. 
For example, this is the case for \tess, \cheops, \hstgsevenfivezero\ and \hstgfourthreezero\ filters for some (but not all) 
models inside the temperature range 1000~K$\geqslant$$\teq$$\leqslant$1400~K. This can be understood by noting that the effect of metallicity is 
to move the light formation depth toward lower pressures. For short wavelength filters (and especially for models without \tio/\vo\ opacity),
the light formation depth in metal rich models can now cover the region there the temperature distribution is more asymmetric compared to models
with solar metallicity (normally around 10$^{-1}$~bar) and thus stronger offsets can be observed. 
Because the temperature distribution varies with planet
orbital period and gravity, the way how short wavelength filters react on changing metallicity appears complicated.
On another hand, long wavelength filters
in solar metallicity case already probe high altitudes where temperature distribution is less affected by the flow. 
Thus, moving the light formation depth toward even lower pressures results in the reduction of the offset values. 
Finally and following arguments given above, the metallicity alone cannot explain the scatter
in observed points presented in our Fig.~\ref{fig:offset-obs}. Indeed, most of the scatter in \spitzer\ wavelengths, for example, comes from inconsistency 
in individual measurements and/or negative offset values. Increasing metallicity would only decrease theoretical offsets and could
not explain negative values.

Contrary to our model predictions, observations do not show a clear dependence of phase-curve offsets
on planet $\teq$. Still, and especially when ignoring observed negative offsets, the models tend to
align closer to the observed trend in such filters such as \tess, \spitzeriracthree, and \spitzeriracfour.
Similar conclusion was strengthened in the study
by \citet{2022AJ....163..256M} who noted an absence of clear correlation between observed offsets and other
planet parameters, while noticing a weak correlation between offset 
and planet gravity (though only for planets around $\teq$=1300~K), and a statistically significant trend of increasing
phase-curve offset with rotation period. We do find same trend in our simulations
where hot planets (and thus those that can have short rotation periods) show smaller offsets compared to cooler ones.

The circulation models that we employed in this study cannot reproduce commonly observed negative offsets. 
This suggests that we lack some important physical process that could render the atmospheric flows in deep layers
to go in retrograde direction. Note that some hydrodynamic models indicate that both eastward and westward hotspot shifts
are possible under certain conditions \citep{2025ApJ...995...84D}.
The interaction of the magnetic field
with atmospheric flows can be considered as a plausible mechanism to trigger retrograde circulations as well
\citep[e.g.,][but notice a rather hot temperature $\teq$$=$2400~K of their model]{2025A&A...699A.339B},
or even lead to oscillation between prograde and retrograde motions \citep{2025ApJ...978..149H,2017NatAs...1E.131R}.
In addition, clouds on the western part of the planetary dayside can result in negative offsets 
specifically at optical wavelengths, too \citep{2016ApJ...828...22P}.

\subsection{Photochemistry, mixing processes, and clouds}

In our simulations we assumed that the atmospheric constituents are in
chemical equilibrium. This assumption is not strictly valid at the sub-stellar point where
the concentrations of atomic and molecular species must be calculated taking into account
energy of the incoming radiation from the parent star. The most affected 
molecules are \hho, \chhhh, \nhhh, and \hh, and they contribute to the major 
bands in the visible and infrared spectra of planets. 
Indeed, some of the photochemical products have been discovered in the past \citep[e.g.,][]{2023Natur.617..483T}.
We did not include photochemistry in our simulations because
such models are computationally expensive for our approach as we would 
need to calculate photochemical processes in radial direction for each surface element.
Moreover, we would need to extend substantially the chemical networks that are currently implemented in
available photochemical codes by including chemical reactions for the refractory elements (Fe, Si, Cr, etc.).
This renders the problem of including photochemistry extremely challenging at the moment.
On the other hand, photochemical process are often taking place in the outermost regions of a planetary
atmosphere with pressures $p$$<$10$^{-2}$~bar, depending on the temperature structure.
For instance, strong temperature inversion layers can act as an effective quenching barrier \citep[see, e.g.,][]{2020A&A...639A..48S}. 
Nevertheless, most significant changes in atmospheric chemistry occur in the layers that do not 
contribute much to the emergent radiation, and thus
simulated phase curves are likely to be less affected by them.
This, however, must be addressed in detail in future studies.

Next, we ignored particle transport processes and the turbulent mixing. 
For instance, molecular diffusion process are of particular importance at lower temperatures
\citep[see, e.g.,][]{2020A&A...639A..48S}, where it could modify the strength of molecular bands, such as 
\chhhh, \nhhh, \hcn\ \citep[see also, e.g.,][]{2024ApJ...963...41T,2006ApJ...649.1048C}.
Strong horizontal winds can also change species concentrations by mixing atmospheric material
between day and night side, and many other 3D phenomena that could potentially
affect the spectra of planets \citep[see, e.g.,][]{2020A&A...635A..31M,2018ApJ...855L..31D}.
Whereas chemical equilibrium is likely a good assumption for the hottest planets, 
the carbon chemistry of the cooler objects will likely be affected by transport processes, 
either vertical or horizontal.

Finally, it is known that some HJs may have thick cloud decks \citep{2016Natur.529...59S}. 
The effect of clouds is to weaken spectroscopic features due to scattering absorption. 
Thus, this effect is largest at short (optical and near-infrared) wavelengths 
where the scattering absorption is the strongest.
The presence of clouds and hazes can mute the photochemically produced features at \tess\ and \cheops\ wavelengths,
and also in many \nircam\ wavelengths, such as between 0.6~\mum\ and 3~\mum.
For instance, a theoretical study by \citet{2021ApJ...908..101R} on phase curves of HJs
showed that the cloud formation may significantly alter the heat redistribution and wind speeds in
planets with $\teq$$<$2300~K. Their simulations did not extend toward multiwavelength phase curves
and were only represented by two broadband wavelength channels, visible and infrared, respectively,
and only reflected light was considered for the visible channel.
A general effect was that, for the cloud-free cases, the phase-curve offset decreased with $\teq$, which
is what we also find in this work. However, in some cases clouds were found to be able to revert this trend 
and to decrease the offset as $\teq$ decreases in their infrared channel. 
In the visible channel clouds were found to produce even retrograde
hotspot offsets in some models, and no significant offsets for coolest temperatures ($\teq$$<$1600~K).
Nevertheless, clouds are not expected to seriously affect spectra of planetary atmospheres at
$\lambda$$>$10~\mum, except perhaps to condensate vibrations mode features, 
which can become prominent in the infrared \citep{2015A&A...573A.122W}.
Also, thick clouds that are formed on the planet nightside will reduce the phase-curve offset and increase the phase-curve amplitude
in all wavelengths where the cloud opacity is relevant \citep[see, e.g.,][]{2021MNRAS.501...78P,2016ApJ...828...22P}.

\subsection{Deviations from the local thermodynamic equilibrium}

Our calculations of planet emission was based on LTE assumption that states that the population of energy levels within atoms and molecules,
as well as the velocity distribution of free electrons, obey Boltzmann-Maxwell distribution. This, however, is not the case at the substellar point
where the radiation from the host can be high enough to dominate rates of level populations.
For instance, \citet{2021A&A...653A..52F,2025A&A...699A.186F} showed that strong
irradiation from the host star drives the population of energy level of many atomic species away from their equilibrium values in very hot planets
KELT-9b and WASP-178b.
It is however unclear what effect would NLTE have in other HJs. 
Indeed, it is not the $\teq$ of the planet that can solely be used as a proxy for the magnitude of NLTE effects,
but, more importantly, the high energy XUV flux of the host star.
For instance, a study by \citet{2023A&A...676A..99F} of KELT-20b found that the relative atmospheric heating due to NLTE effects in this planet
is \textit{stronger} compared to a hotter KELT-9b, for example. What is common for both planets is their host stars that are of spectral type A
and thus have strong XUV flux. 
However, at this moment our models do not account for the NLTE processes and thus we ignored them in the present study.

\subsection{Suggested observing strategy}

In order to constrain flows in HJ atmospheres, observations in a set of wavelengths
that probe physically very different atmospheric altitudes are essential.
Atmospheric flows at high altitudes will be weak due to a low density of the 
planet atmosphere and hence short radiative timescales, while flows at higher pressures
will be strong. This is exactly what we see from the analysis of the phase shifts
where filters that probe high altitudes demonstrate smallest offsets.

Among all instruments that we considered in our study, the \nircam\ appears to be very
promising in studying atmospheric dynamics. It contains filters that probe high and low
altitudes on day and night sides. For instance, a combination of F070W, F090W, F115W, F150W2, and
F460M~--~F480M filters could provide a consistent look into the flows originating 
at various depths, for example, between 10~bar and 10$^{-2}$~bar for cool planets without \tio/\vo, 
and between 1~bar and 10$^{-4}$~bar for hotter ones, respectively.
Another instrument, \miri, also possesses
a set of filters that probe low pressures very well, but the high pressure regions around 1~bar and below
is difficult to probe. In this regard a use of the \nirspec\ Prism appears to be a good choice to study
deep layers at least in dense HJs with $\teq$$<$1600~K. Notice that presently only the \nirspec\ Prism
appears to be observationally feasible choice due to its efficiency which is superior compared to splitting the light
between individual \nirspec\ filters, for example. Still, multiwavelength observations are required if one wants
to extract detailed information about atmospheric flows.

The broadband photometry available from \hst\ also offers a set of filters suitable for probing both high- and low-altitude atmospheric flows.
Our simulations indicated that the best combination
is \hstgfourthreezero\ and \hstwfc, where the former probes high altitudes and the latter probes mostly low altitudes
both on day and night sides, respectively.

\tess\ and \cheops filters are good in probing high altitudes on dayside specifically in HJs with temperatures
around $\teq$$\approx$1600~K. But they lack sensitivity to the deep layers compared to \hstwfc, for example.
The soon coming \plato\ mission\footnote{https://www.esa.int/Science\_Exploration/Space\_Science/Plato} 
(PLAnetary Transits and Oscillations of stars) is expected to show sensitivity to atmospheric flows similar to \tess.

\section{Summary}

In this work we utilized SPARC/MITgcm general circulation models of hot Jupiters to investigate
the possibility of using various wavelength domains to resolve the vertical structure of atmospheric flows. 
To do this we simulated phase curves in the photometric filters of former and present space missions, such as, 
\spitzer, \tess, \cheops, \hst, and \jwst. We also calculated the effective formations depths and contribution functions
to understand the sensitivity of each filter to changes in atmospheric flows.
Our main conclusions are summarized below.

\begin{itemize}
\item
We confirm a clear correlation of the predicted phase-curve offsets with planet equilibrium temperature,
in agreement with recent findings by, for example, \citet{2022AJ....163..256M}.
{We restricted our study to include only solar metallicity models, but note that high metallicity can also play an important role
in reducing phase-curve offsets \citep[see, e.g.,][]{2024MNRAS.531.1056R}.}
The dependence on atmospheric gravity is of secondary importance in most of the cases, except several \nircam\ filters around 2~\mum\ and for
models with $\teq$$\leqslant$1400~K.

\item
Available observations of phase-curve offsets do not show a clear dependence on $\teq$ or other parameters.
The only exception seems to be the \tess\ and \spitzeriracfour\ filters for which the deviation between models and observations seems minimal. 
Still, our models cannot explain the retrograde (westward) offsets seen in \tess\ and some \spitzer\ filters.
This must be addressed by using advanced models that account, for example, for magnetic fields and their interaction with atmospheric flows
\citep{2025A&A...699A.339B,2025ApJ...978..149H} as well as clouds for cooler planets \citep{2021MNRAS.501...78P,2016ApJ...828...22P}, 
which is seen to be a plausible explanation for the observed patterns in phase-curve offsets.

\item
Our simulations show that the following combination of photometric filters are the best for probing atmospheric flows at high and low altitudes 
and on the day and night sides of hot Jupiters: 

\begin{itemize}
\item
for the \nircam, a combination of F070W, F090W, F115W, F150W2, and
F460M-F480M filters should be sufficient to resolve flows originating between 10~bar and 10$^{-4}$~bar depending on planet $\teq$;
\item
the \nirspec~Prism is capable of probing particularly deep layers on both night and day sides;
\item
for the \hst, the best set of filters is \hstgfourthreezero\ and \hstwfc, where the former probes high altitudes and 
the latter probes mostly low altitudes both on day and night sides, respectively.
\end{itemize}

The \miri\ photometry is less sensitive to the deep atmospheric layers compared to \nircam\ and \hst.
We also note that \tess\ and \cheops\ filters are good at probing high altitudes on the dayside for planets with $\teq$ around 1600~K, 
and low altitudes for cooler planets without \tio/\vo. Generally, the \jwst\ photometry is preferred thanks to its superior sensitivity.

\item
Using high-resolution infrared spectroscopy has a strong potential for detecting atmospheric flows at different altitudes,
especially if simultaneous observations at different spectral windows are possible. Presently, instruments such as
\criresplus\ are capable of detecting velocity variation between the position of spectral lines on the order of $\Delta\upsilon$$\approx$1~\kms, 
which is sufficient for studying the atmospheric dynamics from the emission observations.
\end{itemize}

\begin{acknowledgements}
D.S., L.M.L, and L.G. acknowledge support from the Severo Ochoa grant CEX2021-001131-S funded by MCIN/AEI/10.13039/501100011033.
D.S. and L.M.L. also acknowledge financial support from the project PID2021-126365NB-C21(MCI/AEI/FEDER, UE).
D.C. is supported by the LMU-Munich Fraunhofer-Schwarzschild Fellowship and by the Deutsche Forschungsgemeinschaft 
(DFG, German Research Foundation) under Germany's Excellence Strategy -- EXC~2094 -- 390783311. 
This research has made use of the following services: Spanish Virtual Observatory 
project funded by MCIN/AEI/10.13039/501100011033/ through grant PID2020-112949GB-I00;
the Astrophysics Data System, funded by NASA under Cooperative Agreement 80NSSC21M0056; SIMBAD database, CDS, Strasbourg Astronomical Observatory, France.
D.S., W.D., and M.R. acknowledge the support by the DFG priority program SPP 1992 ``Exploring the Diversity of Extrasolar Planets'' (DFG PR 36 24602/41).
\end{acknowledgements}

\bibliographystyle{aa}
\bibliography{biblio}

\begin{appendix}

\onecolumn

\section{Supplementary figures}

\begin{figure*}[ht!]
\centerline{
\includegraphics[width=\hsize]{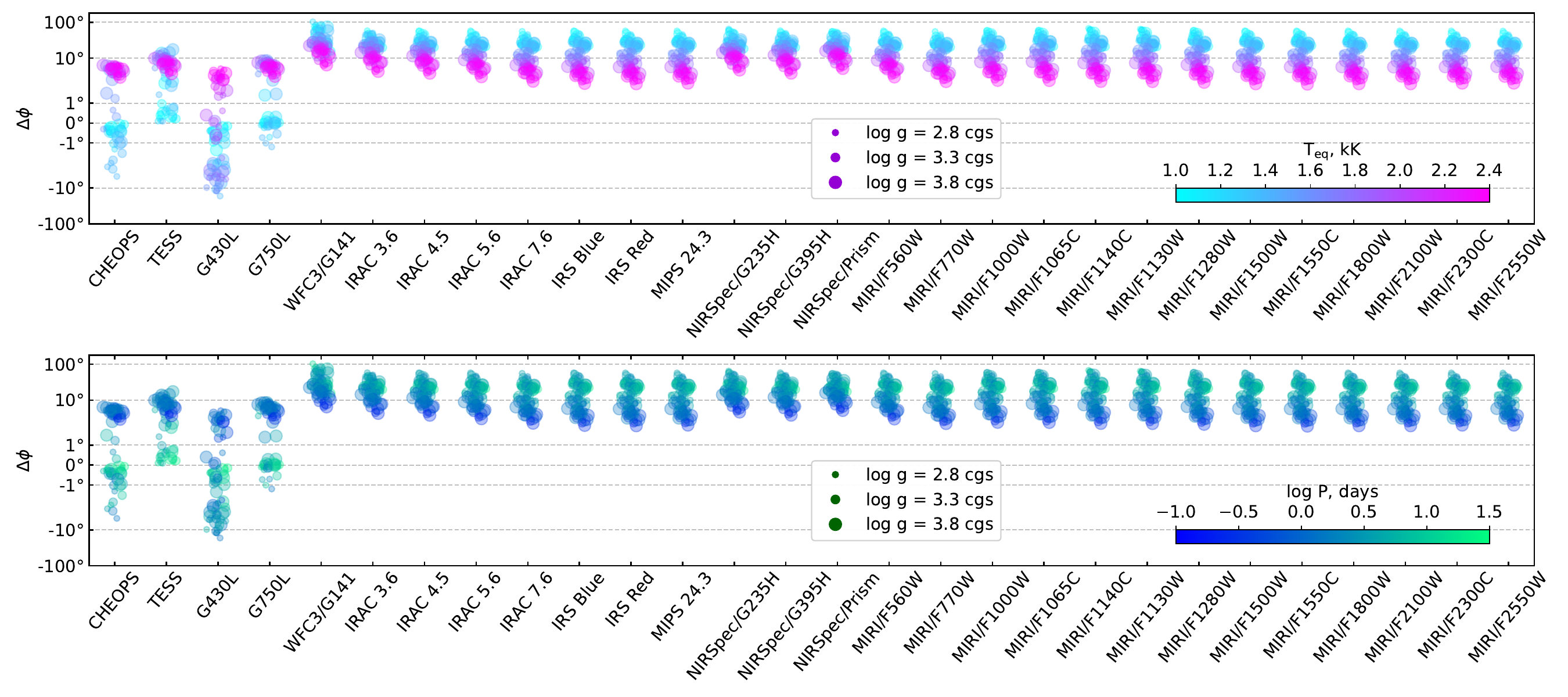}
}
\caption{Predicted phase-curve offsets in \cheops, \tess, \hst, \spitzer, \nirspec, and \miri\ filters plotted in symmetrical logarithm scale.
The symbols are color-coded according to equilibrium temperature (top panel) and rotation period (bottom panel).
The symbol size reflects the $\logg$ of the planet. 
Each point was slightly shifted along x-axis by a random value in order to separate individual models within each filter for a better view.}
\label{fig:offset-filters-set-1-symlog}
\end{figure*}

\begin{figure*}[ht!]
\centerline{
\includegraphics[width=\hsize]{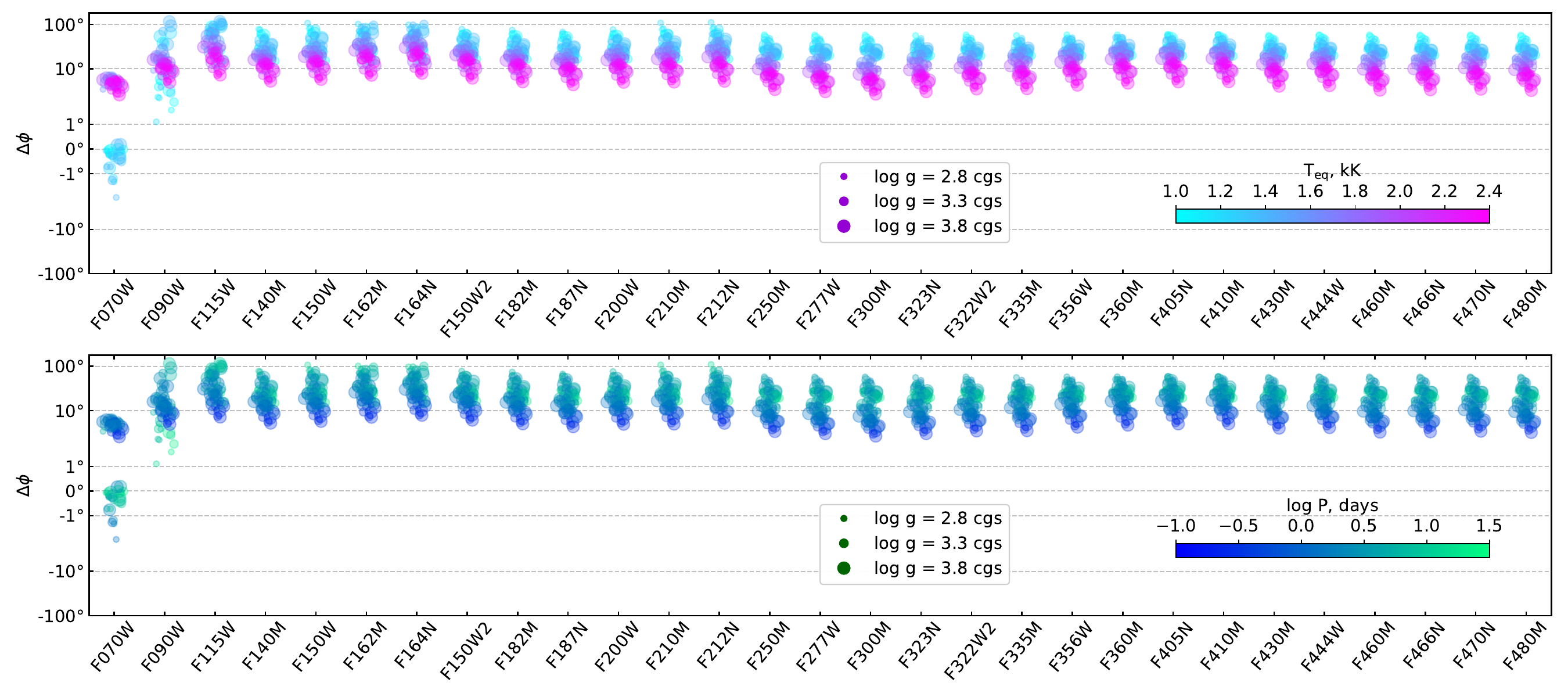}
}
\caption{Same as on Fig.~\ref{fig:offset-filters-set-1-symlog} but for \nircam\ filters.}
\label{fig:offset-filters-set-2-symlog}
\end{figure*}

\FloatBarrier

\begin{figure*}[ht!]
\centerline{
\includegraphics[width=\hsize]{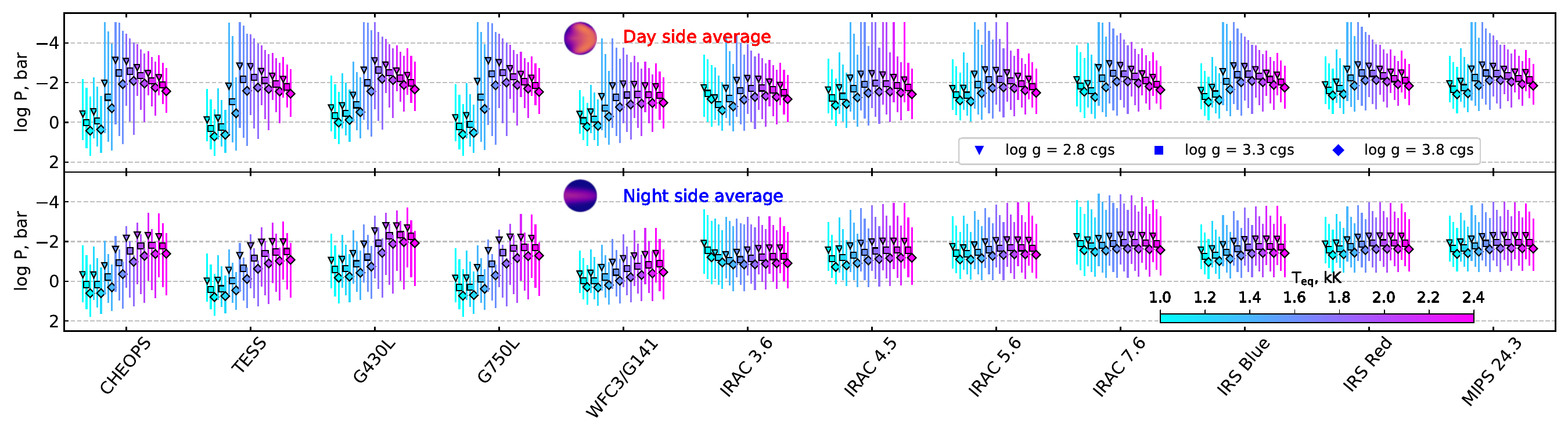}
}
\caption{Light formation depths expressed as effective pressure, $\logpeff$, for the \tess, \cheops, \hst, and \spitzer\ filters. Within each filter,
the predicted data points were shifted along x-axis to separate individual models for better view. The color coding is according to the $\teq$ of the planets while
different symbols refer to the planetary $\logg$ (see the figure legend). Each value of $\logpeff$ is an average over the dayside 
(0.25$\leqslant$$\phi$$<$0.75, top panel) and the nightside (0.75$\leqslant$$\phi$$<$0.25, bottom panel), respectively.
The vertical error bars mark minimum and maximum atmospheric pressures where 
the average value of the contribution function is greater than some arbitrary value (which we chose to be 5\%\ of the peak value)
and hence roughly represents the region of atmospheric depths probed by each filter.}
\label{fig:cf-tess-cheops-hst-spitzer}
\end{figure*}

\begin{figure*}[ht!]
\centerline{
\includegraphics[width=\hsize]{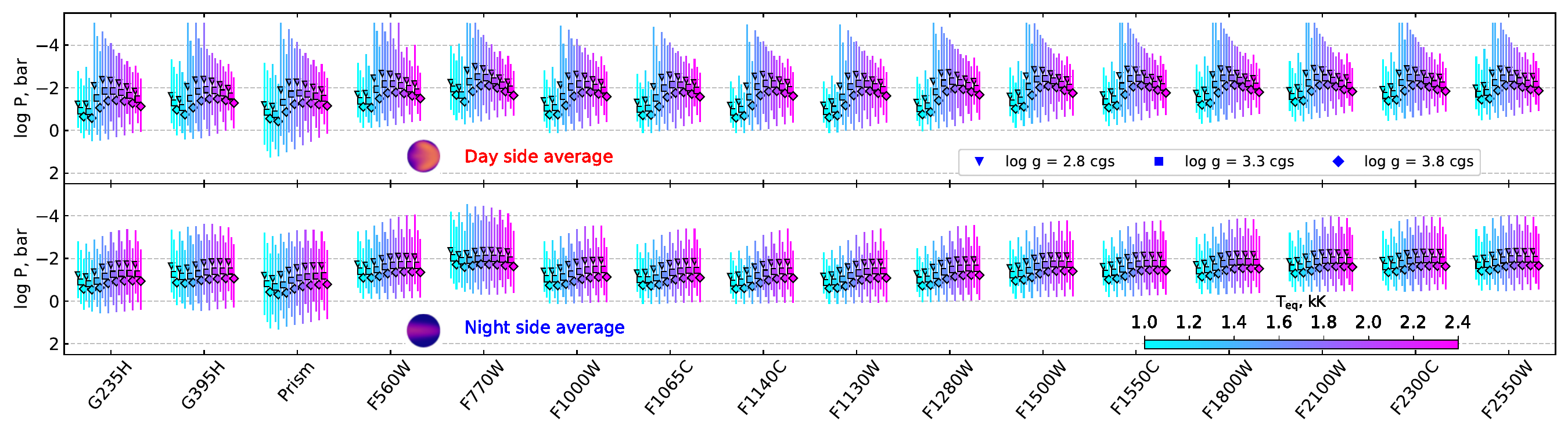}
}
\caption{Same as in Fig.~\ref{fig:cf-tess-cheops-hst-spitzer}, but for \nirspec\ and \miri\ filters.}
\label{fig:cf-nirspec-miri}
\end{figure*}

\begin{figure*}[ht!]
\centerline{
\includegraphics[width=\hsize]{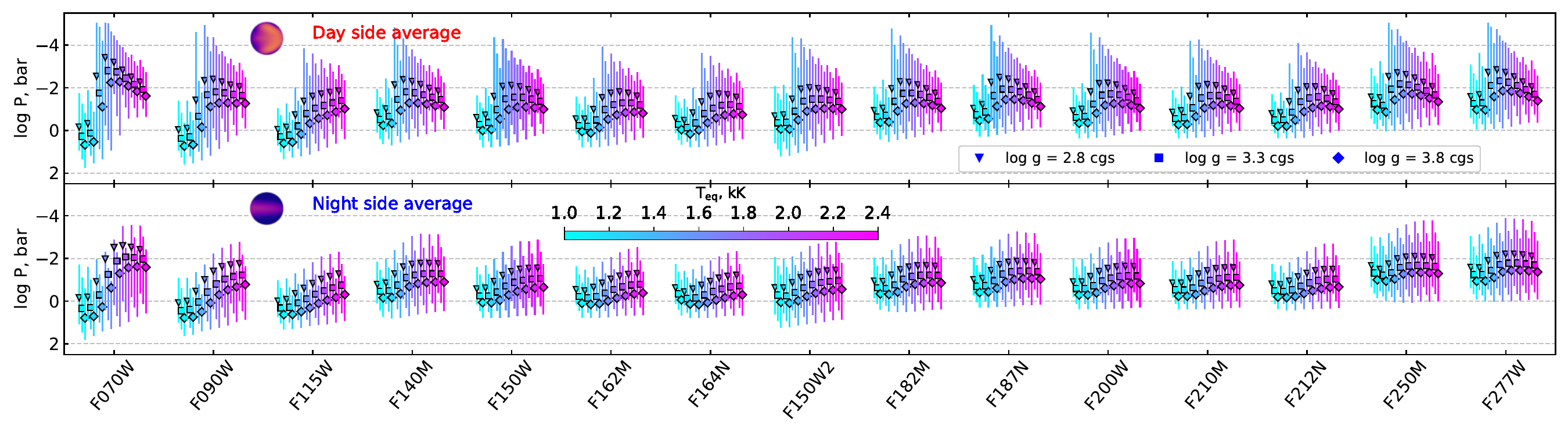}
}
\caption{Same as in Fig.~\ref{fig:cf-tess-cheops-hst-spitzer}, but for \nircam\ filters with effective wavelengths $\lambda_{\rm eff}$$<$3\mum.}
\label{fig:cf-nircam1}
\end{figure*}

\begin{figure*}[ht!]
\centerline{
\includegraphics[width=\hsize]{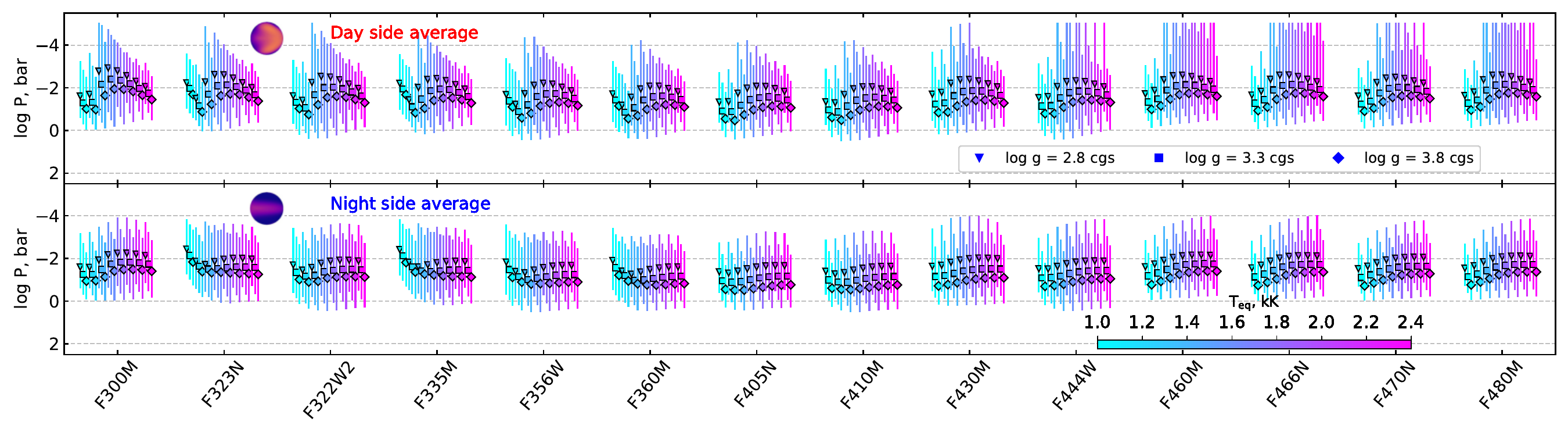}
}
\caption{Same as in Fig.~\ref{fig:cf-tess-cheops-hst-spitzer}, but for \nircam\ filters with effective wavelengths $\lambda_{\rm eff}$$>$3\mum.}
\label{fig:cf-nircam2}
\end{figure*}

\FloatBarrier

\begin{figure}[ht!]
\centerline{
\includegraphics[width=0.7\hsize]{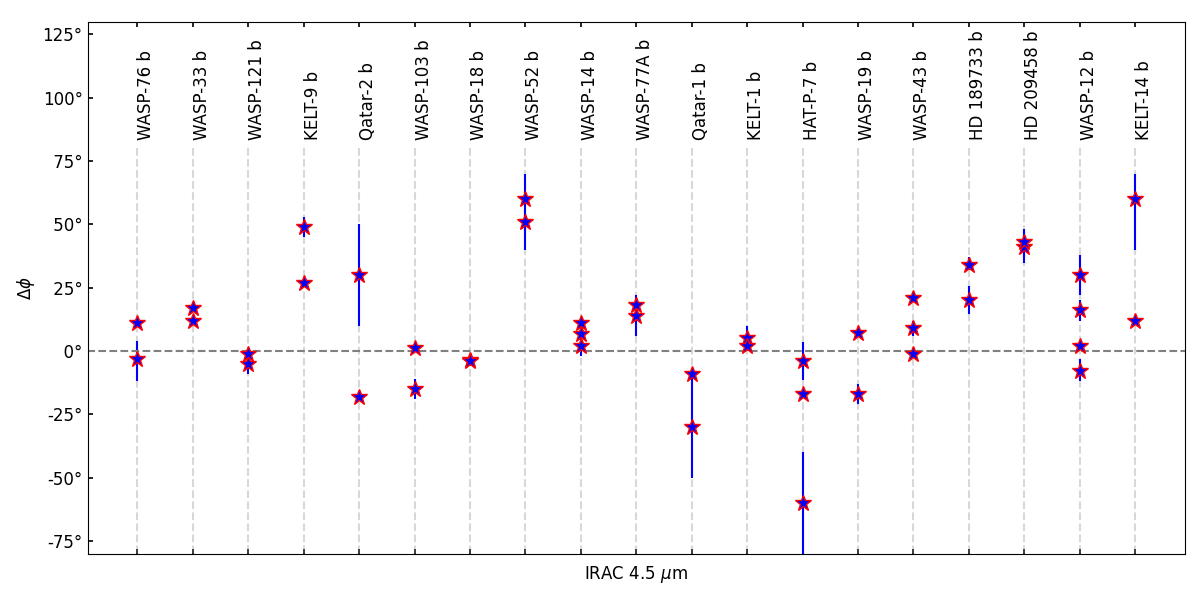}
}
\caption{Example of individual measurements of phase-curve offsets in \spitzeriracfour\ filter for a set of hot Jupiters.}
\label{fig:offset-obs-irac4.4}
\end{figure}

\FloatBarrier

\begin{figure*}[ht!]
\centerline{
\includegraphics[width=0.85\hsize]{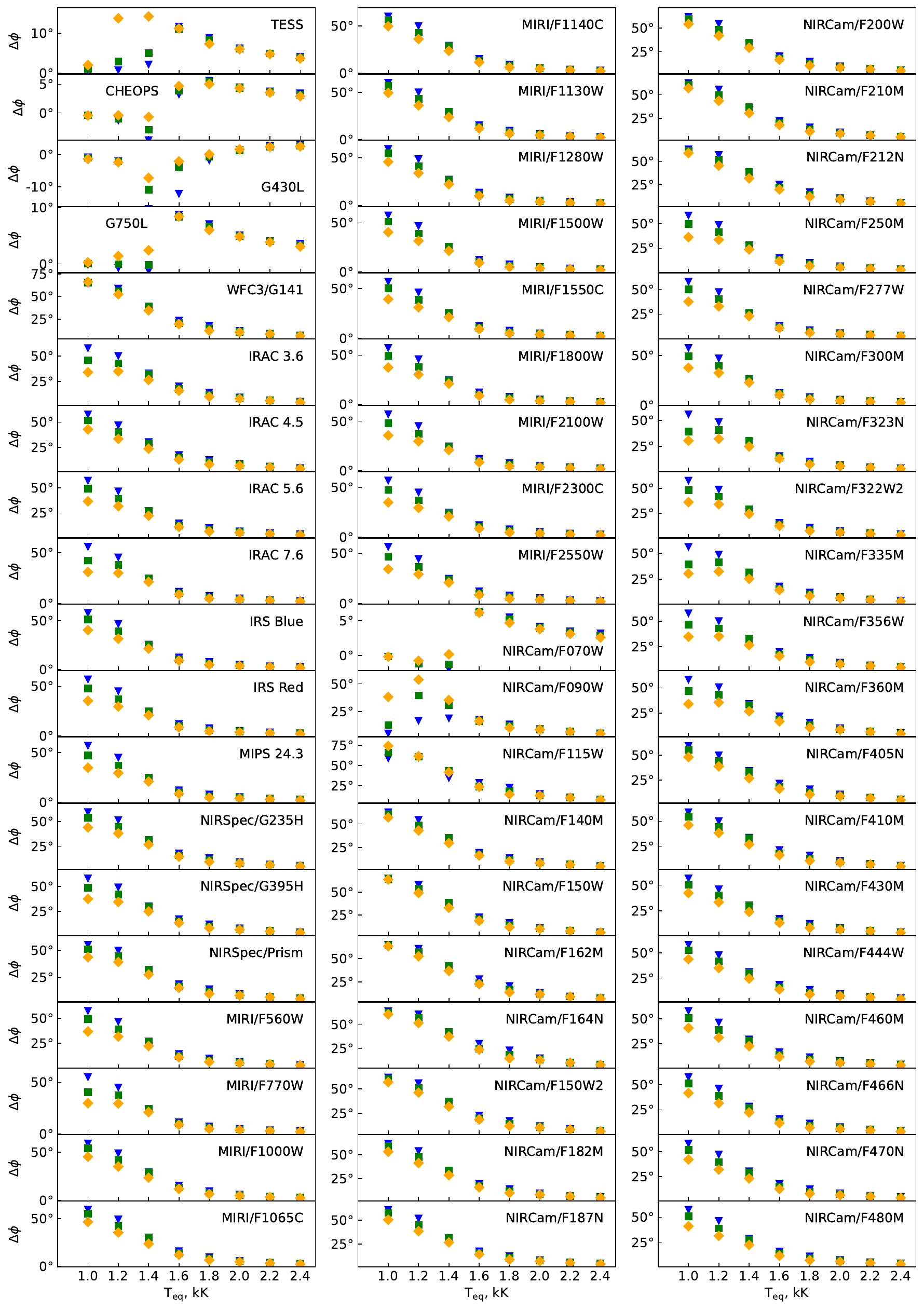}
}
\caption{Predicted phase curves offsets as a function of $\teq$ and gravity for planets orbiting $\mstar$$=$0.8$\Msun$ star.
Different symbols correspond to models with $\logg$$=$2.8~cgs (blue triangles), 
$\logg$$=$3.3~cgs (green squares), and $\logg$$=$3.8~cgs (orange diamonds), respectively.
Note that y-scale can differ between the plots.}
\label{fig:offset-teq-logg-ms0.8}
\end{figure*}

\begin{figure*}[ht!]
\centerline{
\includegraphics[width=0.85\hsize]{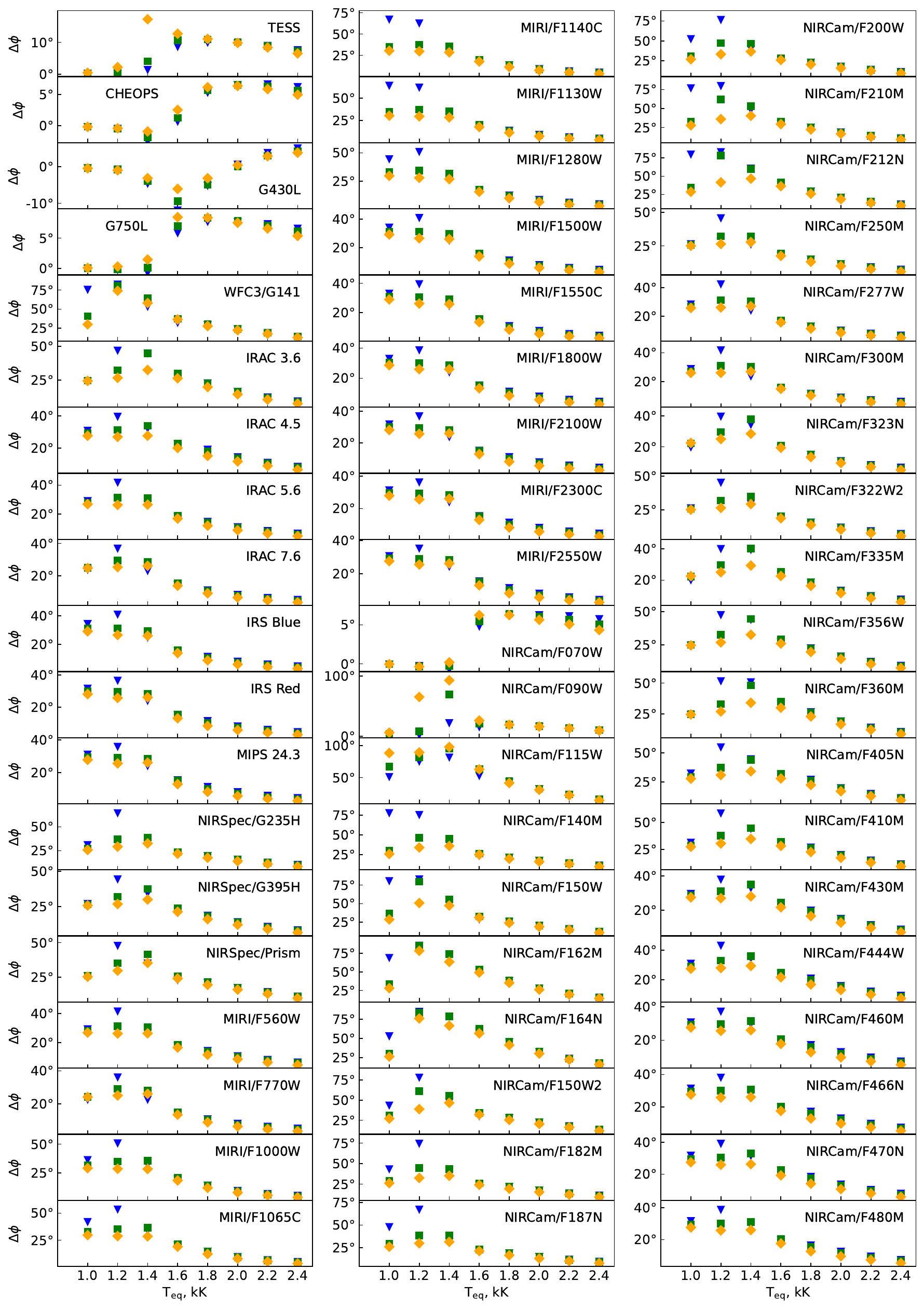}
}
\caption{Same as on Fig.~\ref{fig:offset-teq-logg-ms0.8} but for the planets orbiting $\mstar$$=$1.1$\Msun$ star.
Note that y-scale can differ between the plots.
}
\label{fig:offset-teq-logg-ms1.1}
\end{figure*}

\begin{figure*}[ht!]
\centerline{
\includegraphics[width=0.85\hsize]{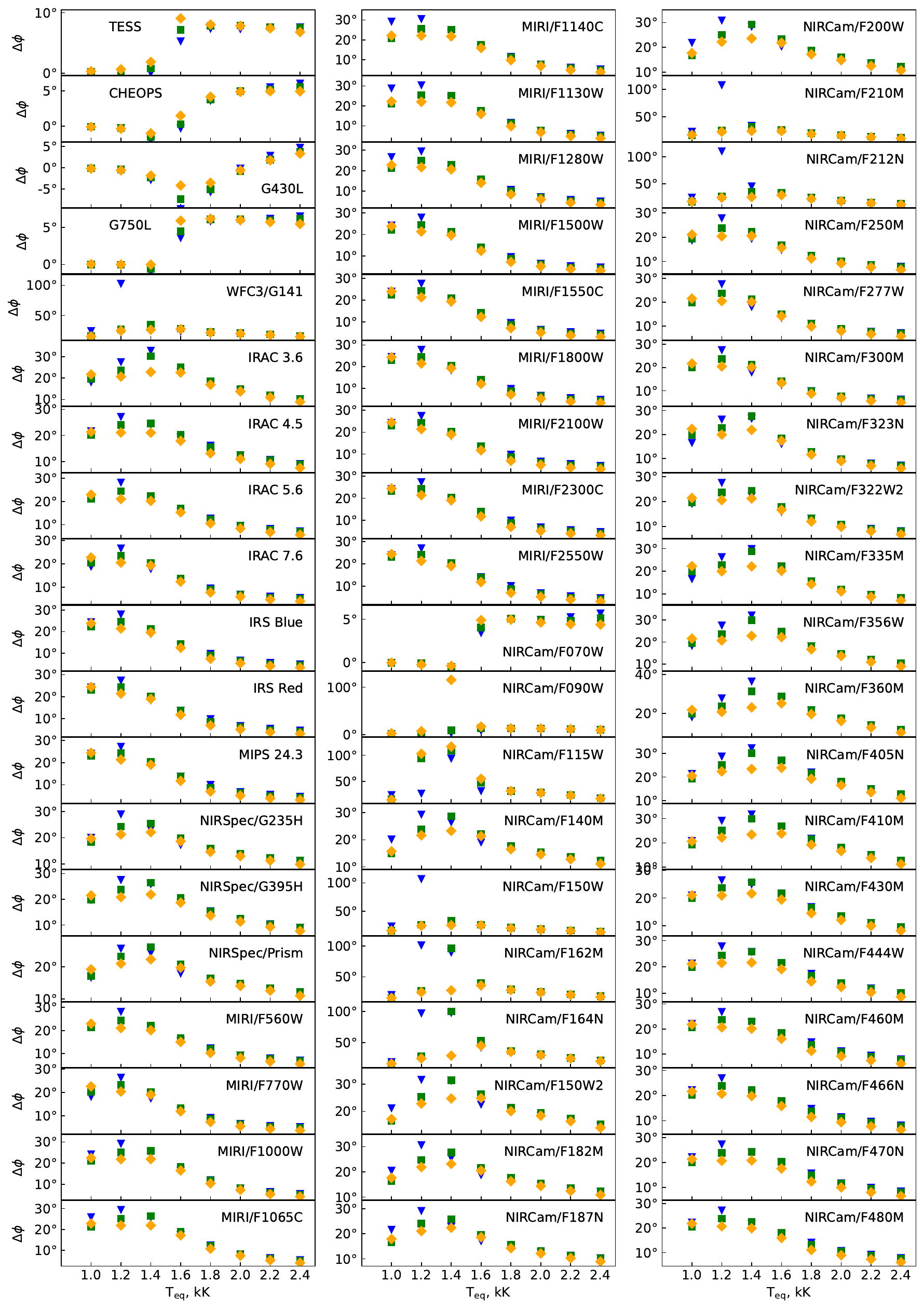}
}
\caption{Same as on Fig.~\ref{fig:offset-teq-logg-ms0.8} but for the planets orbiting $\mstar$$=$1.5$\Msun$ star.
Note that y-scale can differ between the plots.
}
\label{fig:offset-teq-logg-ms1.5}
\end{figure*}

\FloatBarrier

\begin{figure*}
\centerline{
\includegraphics[width=0.5\hsize]{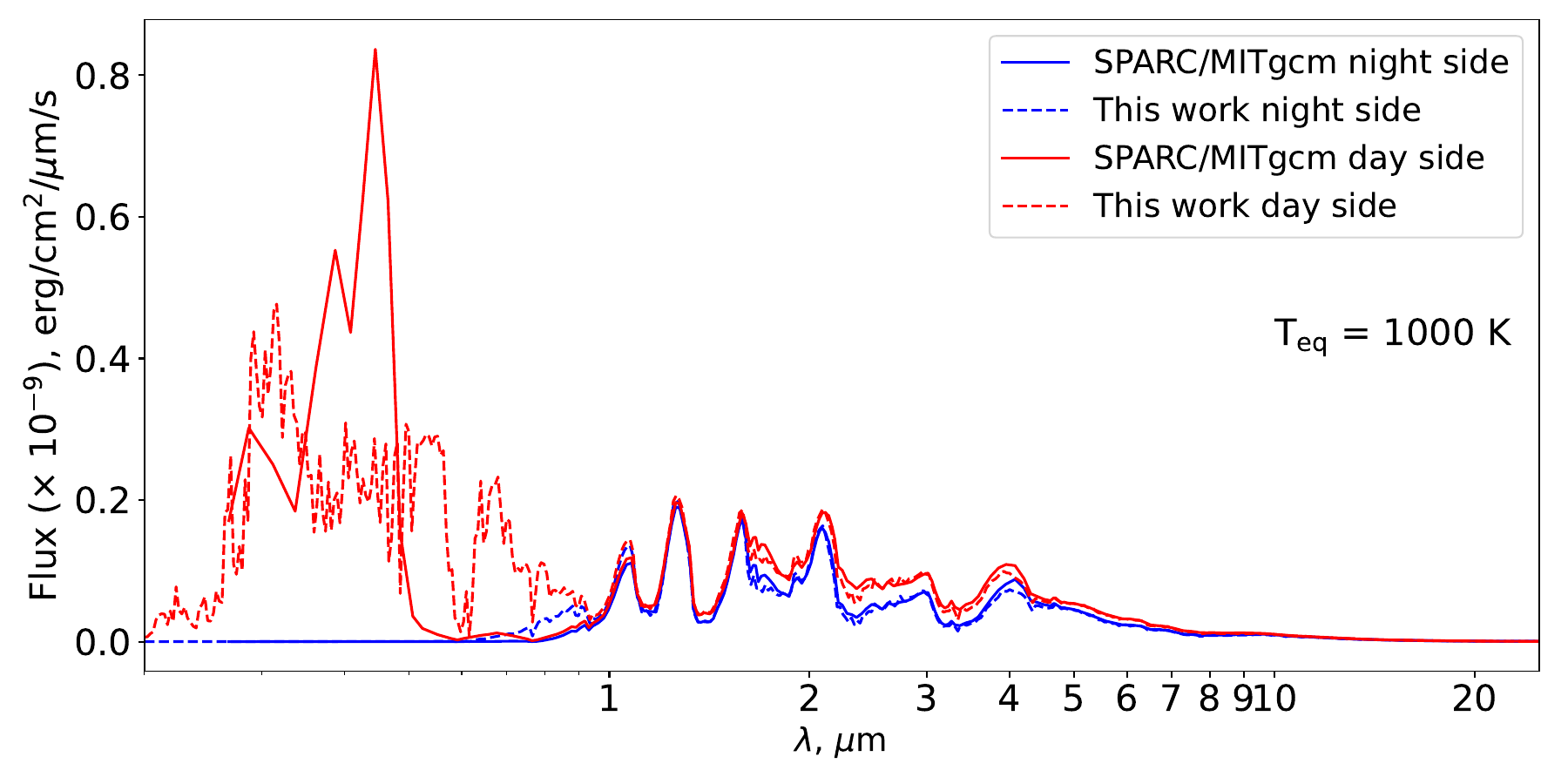}
\includegraphics[width=0.5\hsize]{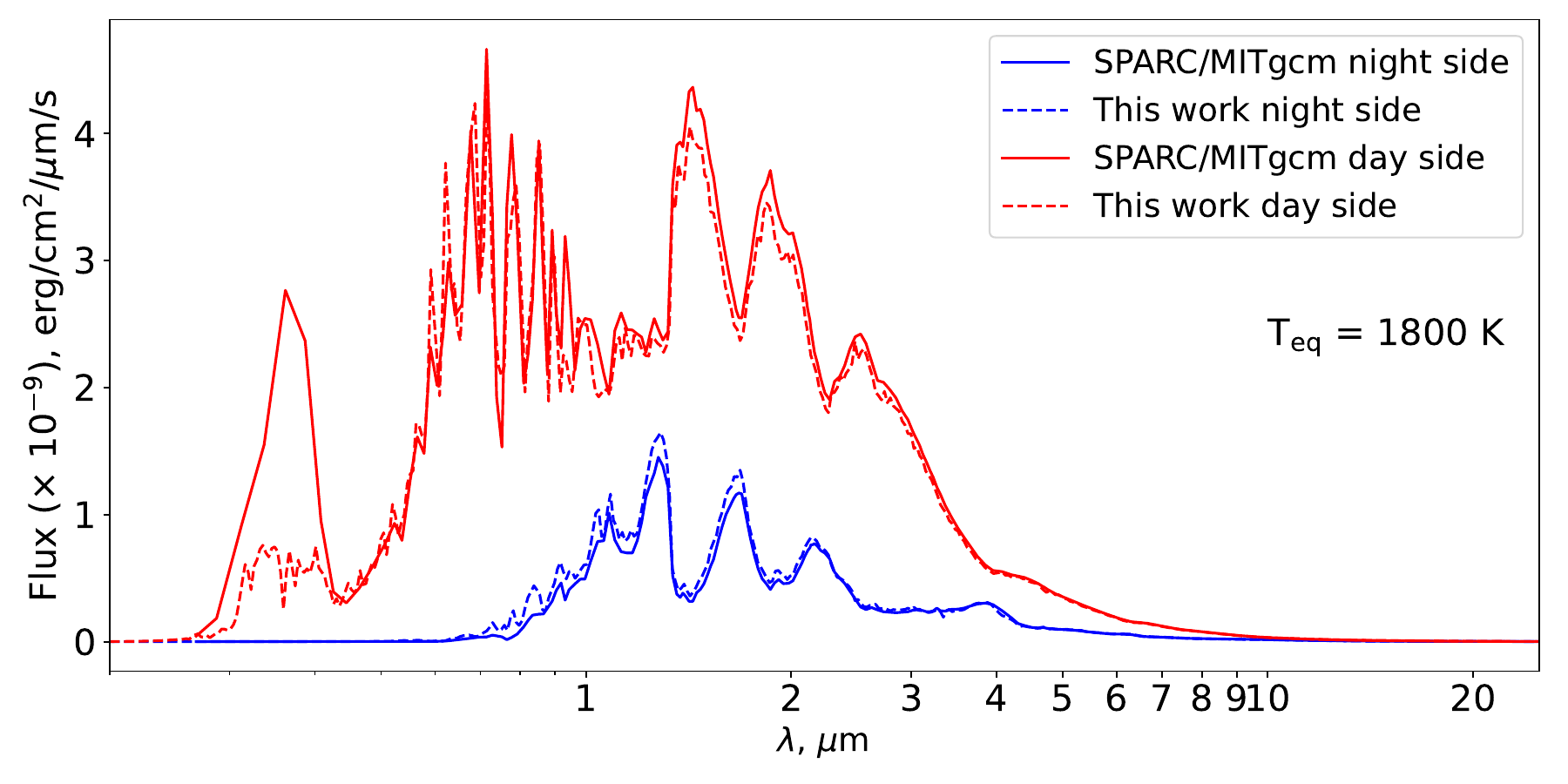}
}
\centerline{
\includegraphics[width=0.5\hsize]{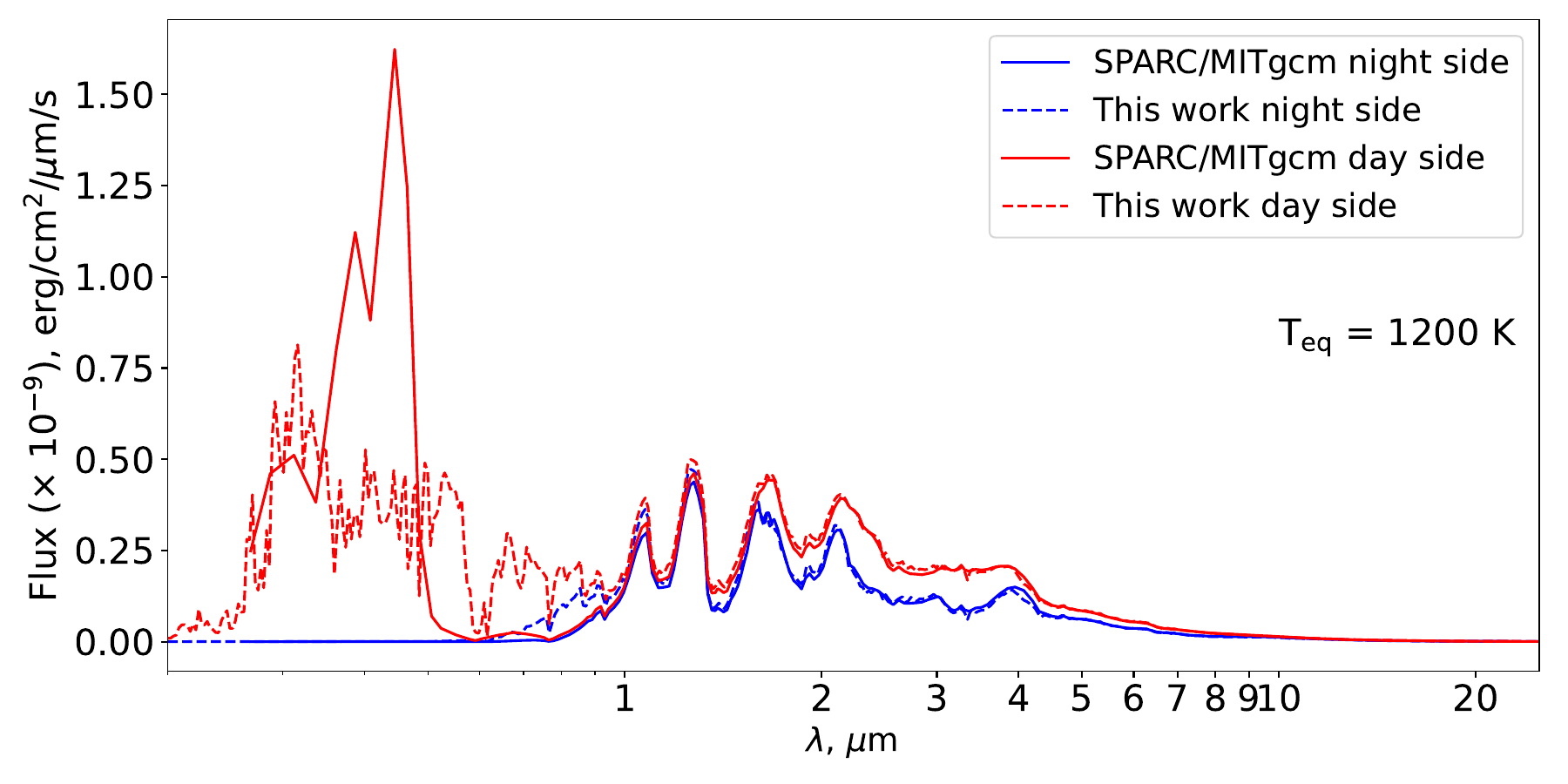}
\includegraphics[width=0.5\hsize]{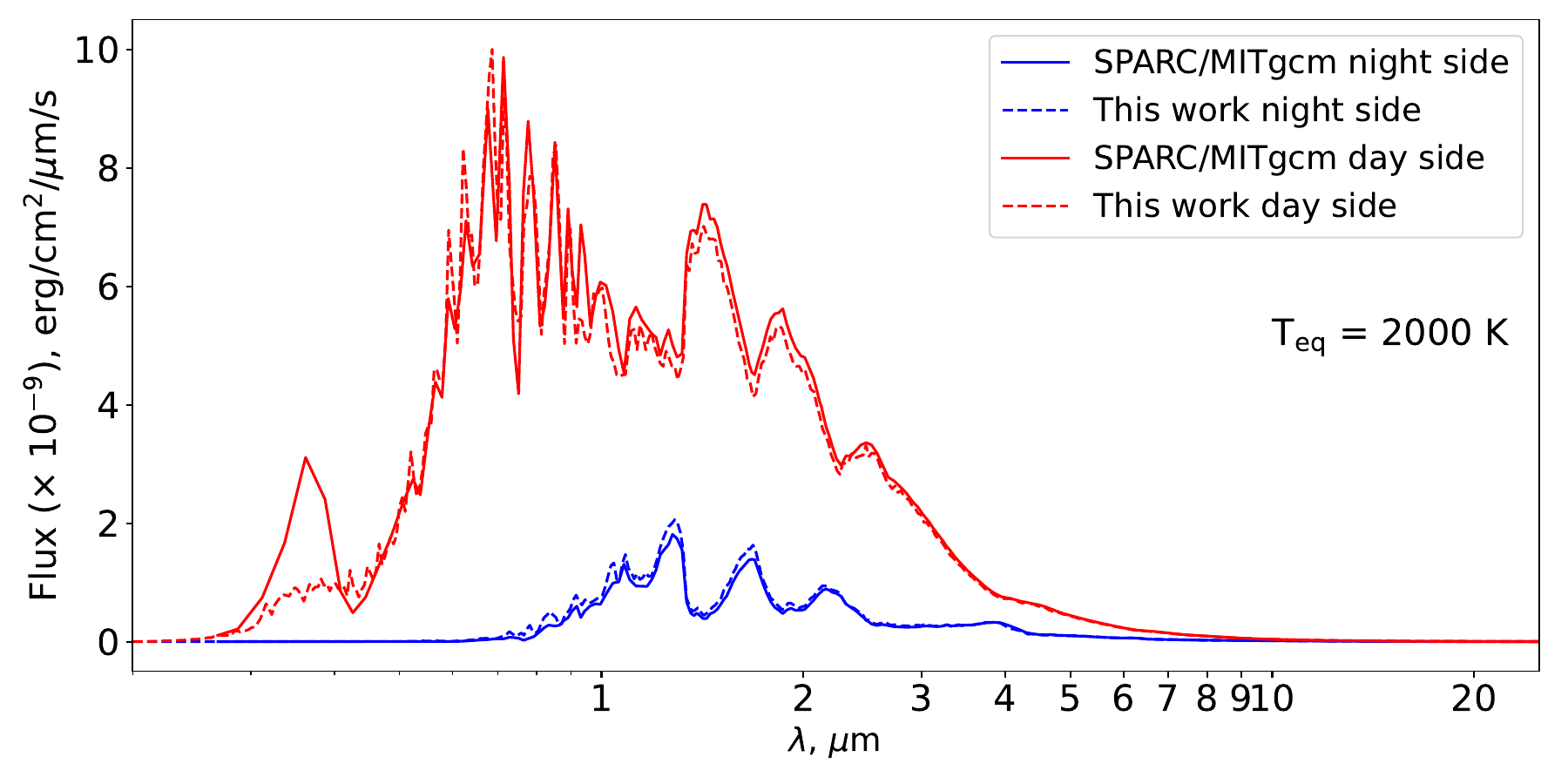}
}
\centerline{
\includegraphics[width=0.5\hsize]{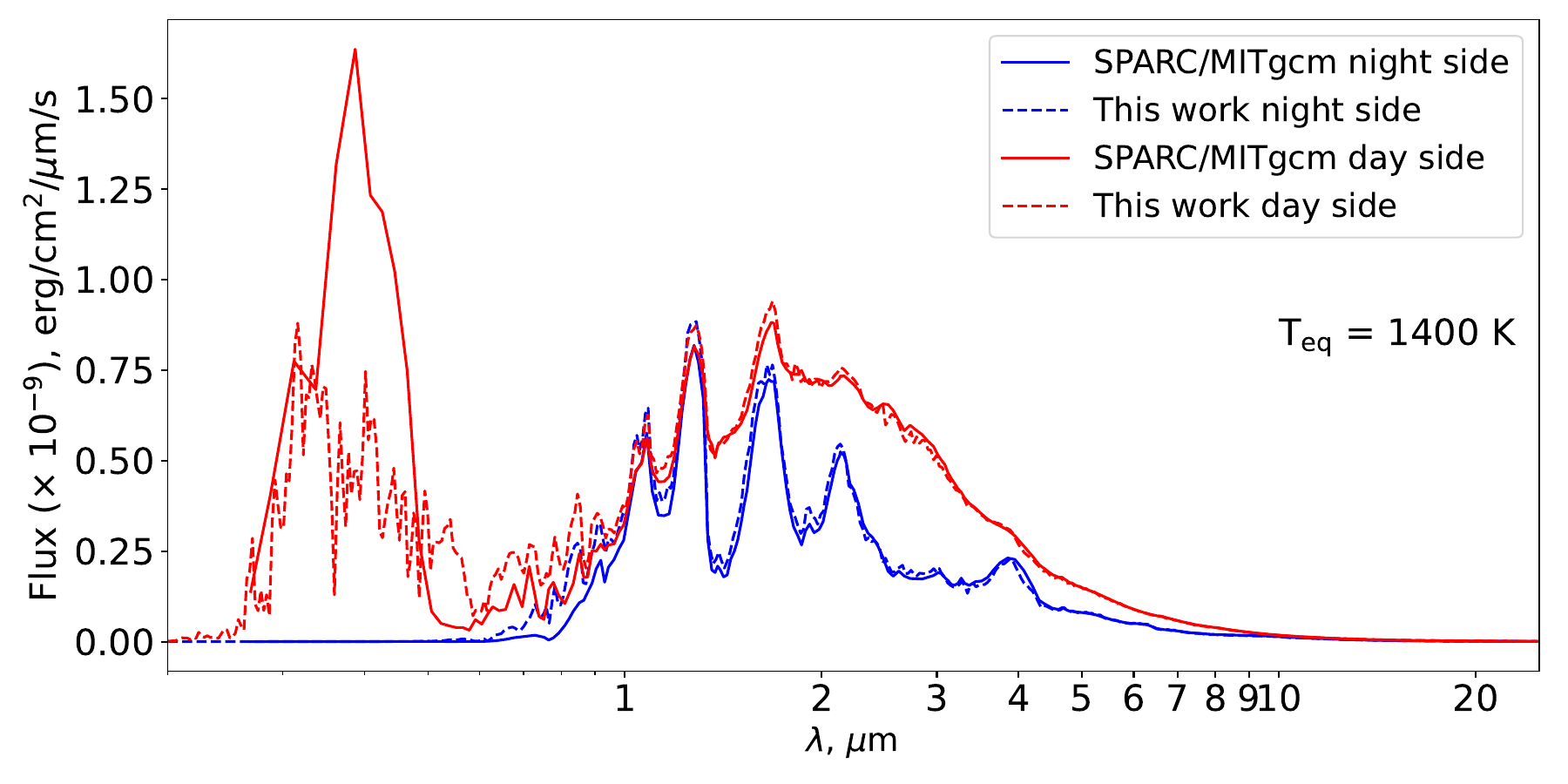}
\includegraphics[width=0.5\hsize]{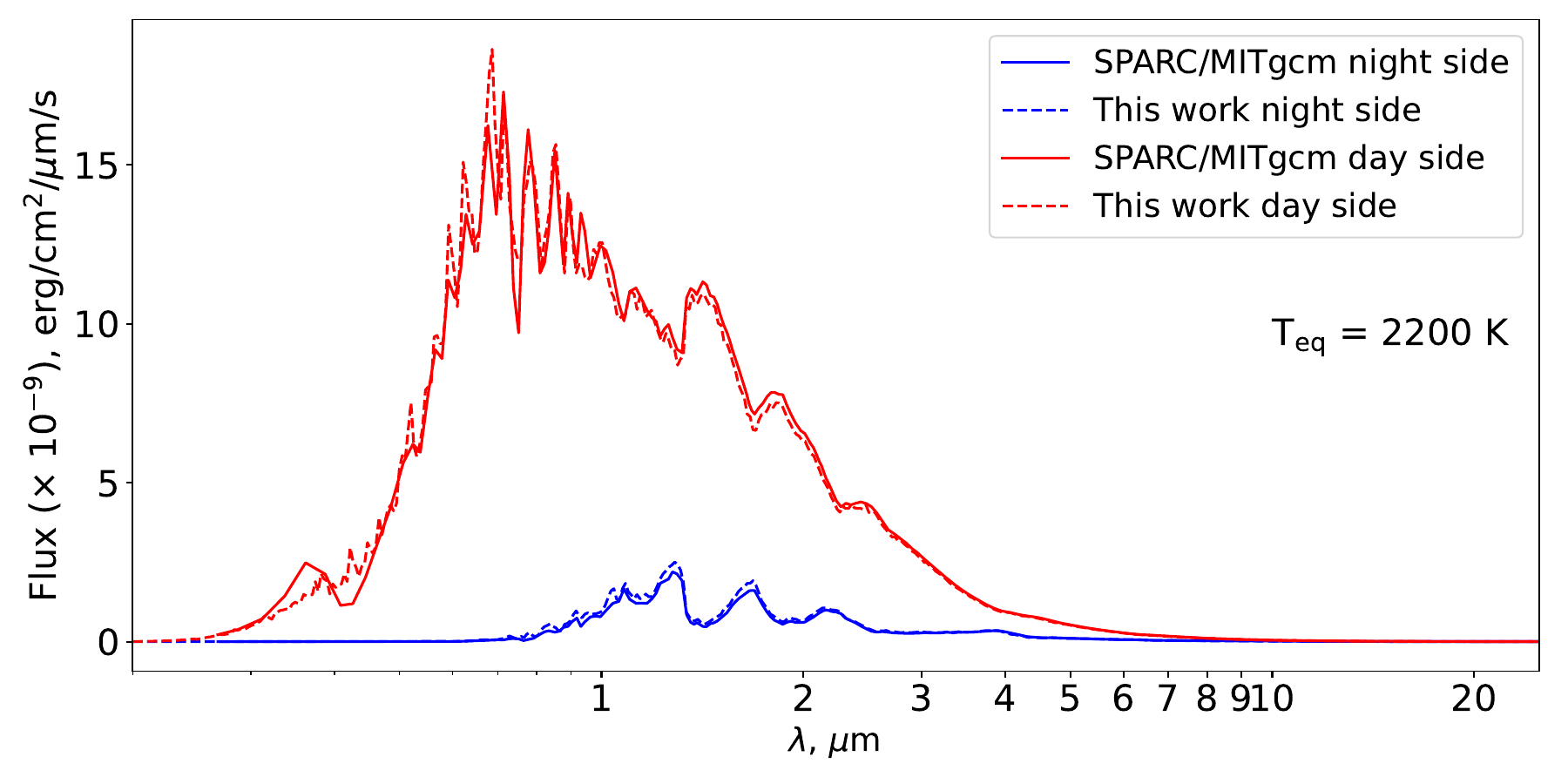}
}
\centerline{
\includegraphics[width=0.5\hsize]{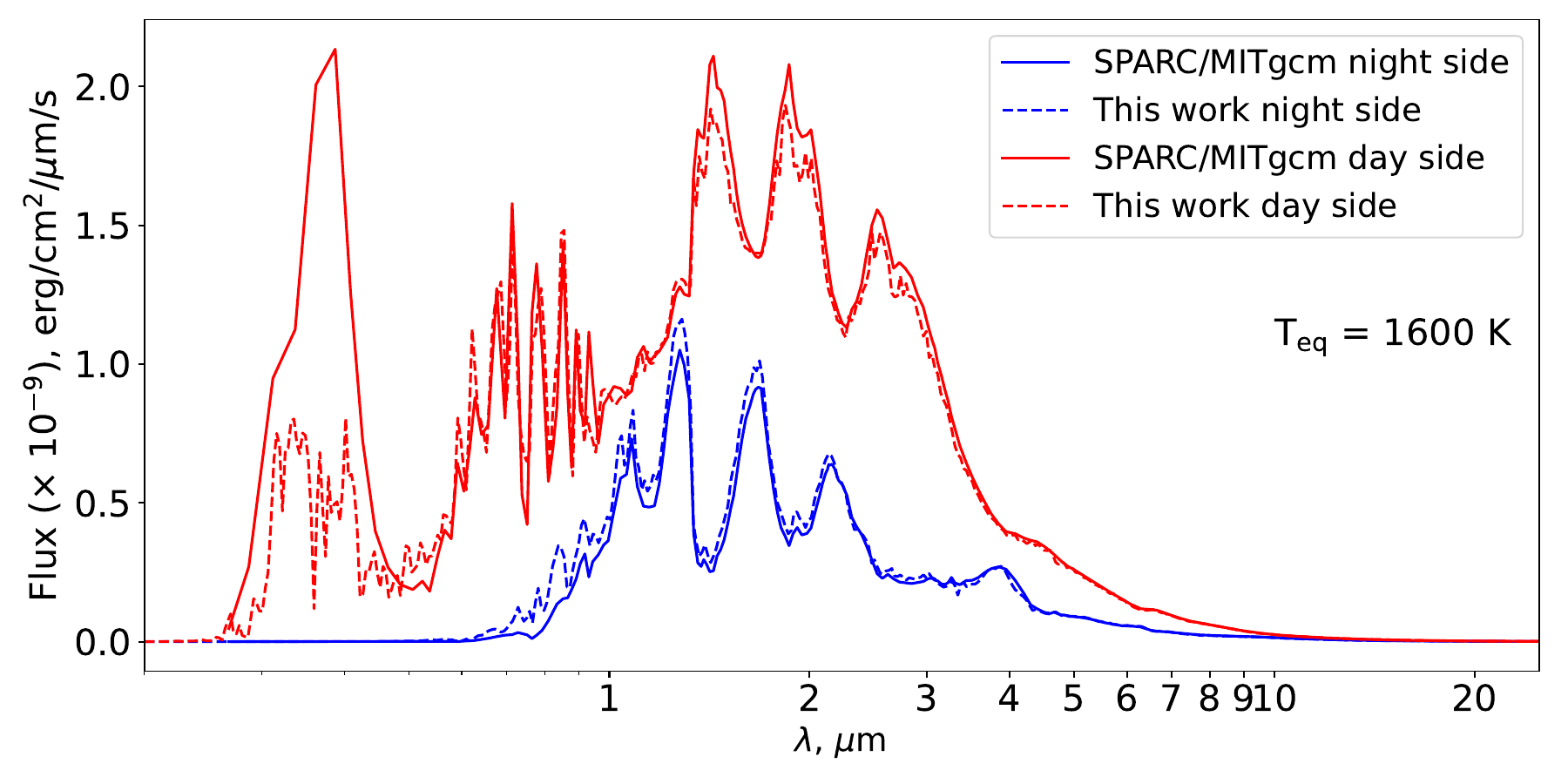}
\includegraphics[width=0.5\hsize]{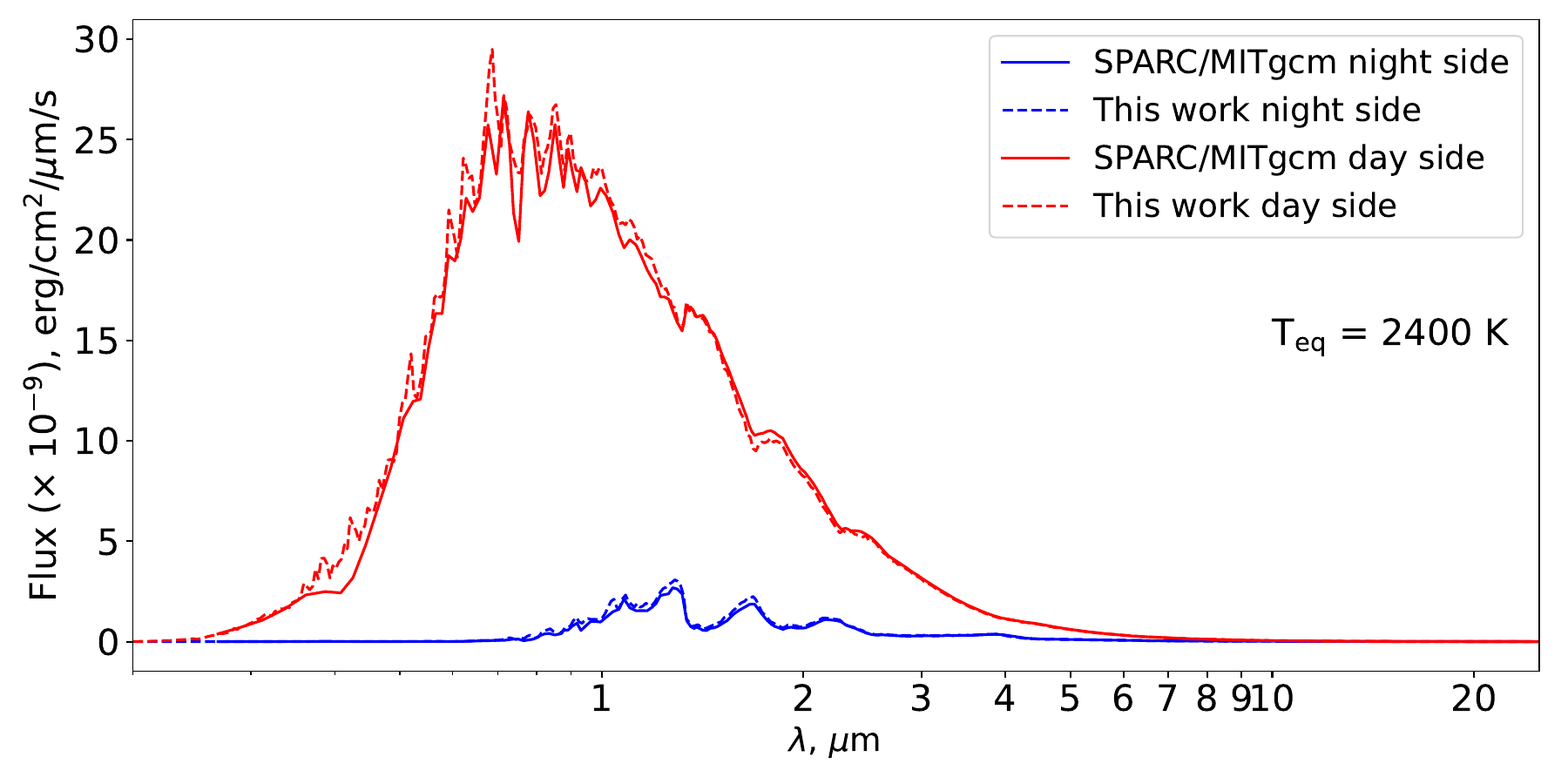}
}
\caption{Comparison between theoretical emergent flux predicted by ADAM (SPARC/MITgcm) from the grid
by \citet{2024MNRAS.531.1056R} and this study for the models having $\logg$$=$3.3~dex, $\mstar$$=$1.1~$\Msun$
but various $\teq$. Shown are the disk integrated flux at orbital
phases $\phi$$=$0.0 (nightside) and $\phi$$=$0.5 (dayside), respectively. See the figure legend for
the description of each curve.
}
\label{fig:flux-teq}
\end{figure*}

\twocolumn

\end{appendix}

\end{document}